\documentclass[english,column, pra,aps,superscriptaddress]{revtex4-1}
\usepackage[T1]{fontenc}
\usepackage[latin9]{inputenc}
\usepackage{color}
\usepackage{textcomp}
\usepackage{amsmath}
\usepackage{graphicx}
\usepackage{babel}
\begin{document}

\title{Entanglement concentration with different measurement in a 3-mode
optomechanical system }

\author{Zhe Li }
\email[]{1219082160@qq.com}

\address{CAS Key Laboratory of Theoretical Physics, Institute of Theoretical Physics, \\Chinese Academy of Sciences, Beijing 100190, China}

 \affiliation{School of Physics, University of Chinese Academy of Sciences,\\  P.O. Box 4588, Beijing 100049, China}

\begin{abstract}
In this work, we perform a series of phonon counting measurement with
different methods in a 3-mode optomechanical system, and we compare
the difference of the entanglement after measurement. In this article
we focus on the three cases: prefect measurement, imperfect measurement and on-off measurement.
We find that whatever measurement you take, the entanglement will
increase. The size of entanglement enhancement is the largest in the perfect
measurement, second in the imperfect measurement, and it is
not obvious in the on-off measurement. We are sure that the more precise
measurement information, the larger entanglement concentration.
\end{abstract}
\maketitle

\section{introductions}

In recent years, quantum entanglement \cite{RH-RMP-2009} has been
regarded as a key source in the quantum information processing, for
it can apply in terms of quantum cryptography \cite{AKE-PRL-1991,TJ-PRL-2000},
allow the realization of quantum teleportation \cite{DB-Nature-1997,CHB-PRL-1993}
and quantum dense coding \cite{CHB-PRL-1992}. A number of strategies
to generate entanglement have been developed in different quantum
systems, such as trapped ions \cite{KM-PRL-1999}, cold atoms \cite{JQY-PRB-2003}
and solid-state qubit \cite{YDW-PRB-2010}. The conventional methods
for entanglement photons rely on nonlinear optical process like
parameter amplification and second harmonic generation. However, the
photons with vastly different frequencies, i.e., microwave photons
and optical photons, can not be entangled directly in this way. Nevertheless,
optomechanical system \cite{MA-RMP-2014} provides probability to
work out this difficulty and some related work have already been done
\cite{YDW-PRL-2012,LT-PRL-2012,YDW-NJP-2012,YDW-PRL-2013,SB-PRL-2012,LT-PRL-2013,MCK-PRA-2013,YDW-PRA-2015}.
An optomechanical system suitable to this purpose is based on an optical
cavity and a microwave cavity interacting with a mechanical element
and such a 3-mode optomechanical system has been realized experimentally
\cite{RA-Nature-2014,CAR-PCS-2011}. 

Entanglement will be severely degraded by the channel noise due to
it's fragile nature. In order to overcome that decoherence effect,
entanglement concentration or entanglement distillation will be utilized.
The idea of the standard entanglement concentration is to exact a
smaller number of elements with higher entanglement by distillation
from a large number of elements with lower entanglement through local
operations and classical communication. From 1996 when Bennett et
al \cite{CHB-PRL-1996} proposed entanglement concentration protocol
firstly to the present, the various entanglement concentration protocol
for discrete-variable \cite{DD-PRL-1996,YBS-CSB-2014} and continuous-variable
\cite{LMD-PRL-2000,JE-PRL-2002} quantum system have been developed
as well as entanglement concentrations have been demonstrated experimentally
\cite{ZZ-PRL-2003,TY-Nature-2003}. \textcolor{black}{However, distilling
continuous-variable entanglement appears to be significantly harder
to achieve than distilling discrete-variable, for one can not distill
a Gaussian state by using only Gaussian operations \cite{JE-PRL-2002,GG-PRA-2002,JF-PRA-2002}.
}Thus non-Gaussian operations, in particular photon counting measurement
\cite{AK-PRA-2006}, are indispensable for Gaussian entangled states
distillation. The photon subtraction strategy, one of the available
experimental operations beyond the Gaussian regime, is based on this
idea. And the non-Gaussian operation with photon counting measurement can be implemented by
beam splitters \cite{TO-PRA-2000,PTC-PRA-2002,DEB-PRA-2003}. 

These ideas motivate us to explore an entanglement concentration protocol 
based on phonon counting measurement for 3-mode optomechanical system
(Fig. 1). In this system, a genuine tripartite entanglement state,
where the two cavity output mode and the mechanical output mode are
entangled with each other, can be generated \cite{YDW-PRA-2015}.
We perform the phonon counting measurement in the mechanical mode
(indirectly through auxiliary photon counting) for the genuine tripartite
entanglement state with different methods. In previous work \cite{WM-SCPMA-2015},
the perfect measurement, i.e., projective measurement, have been considered,
but in practice it is difficult  to find a measurement device which
completely satisfy projective measurement. In this paper, we mainly
focus on and get the general result with imperfect measurement and
on-off measurement, which is available experimentally at present.
While the amount of entanglement after measurement is measured in
terms of logarithmic negativity\cite{GV-PRA-2002}. 
Numerical result and analytical result show that: 1, whatever measure
you take, the entanglement will increase; 2, the entanglement enhancement
is largest in perfect measurement, while smaller enhancement in imperfect
measurement, and it is not obvious in on-off measurement. 3, we are sure that 
the more precise measurement information, the larger entanglement concentration.

The remainder of this paper is organized as follows. In Sec. II, we
introduce the physical system and derive the amount of entanglement
before concentration, along with the generating of a genuine tripartite
entanglement state. In Section III, the the definition of logarithmic
negativity is briefly summarized and the amount of entanglement after
perfect measurement is introduced. Section IV is devoted to the entanglement
distillation with imperfect measurement. We calculated the amount
of entanglement after imperfect measurement perturbative order by
order and compare the entanglement concentration effect analytically.
In Sec. V, we discuss the on-off measurement and derive the average entanglement
after on-off measurement. Finally, we conclude with a discussion and
summary about three different measurement strategies numerically and
analytically in Sec. VI.

\section{physical system and measurement operator}

We consider a three-mode optomechanical system: two cavity modes ($\omega_{1}$
and $\omega_{2}$) are coupled to a single mechanical mode $\omega_{\mathbf{m}}$
(see Fig. 1). The cavities interact with the mechanics via the radiation
pressure \cite{MA-RMP-2014}. The Hamiltonian of the system can be
described by

\begin{figure}
\includegraphics[width=8cm]{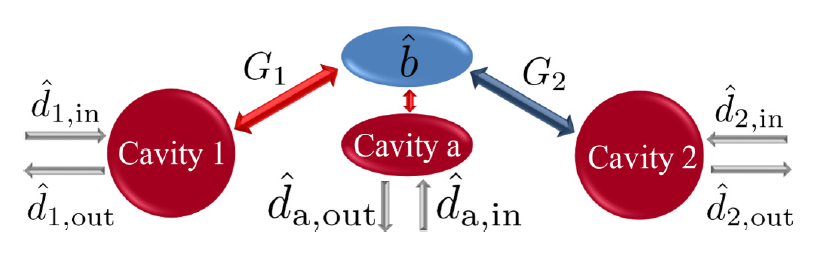}\caption{System schematics. Two driven cavities (cavities 1 and 2) interact
with a common mechanical resonator (mode $\hat{b}$ ). This can generate
entanglement in the optical outputs. An auxiliary third cavity (cavity
a) can be used to cavity cool the mechanics and to make the mechanical
output mode accessible}
\end{figure}

\begin{equation}
\hat{H}=\omega_{\mathbf{m}}\hat{b}^{\dagger}\hat{b}+\sum_{j=1,2}\left[\omega_{j}\hat{a}_{j}^{\dagger}\hat{a}_{j}+g_{j}\left(\hat{b}^{\dagger}+\hat{b}\right)\hat{a}_{j}^{\dagger}\hat{a}_{j}\right],
\end{equation}
where $\hat{a}_{j}$ and $\hat{b}$ are the annihilation operator
for cavity $j(j=1,2)$ and the mechanical mode respectively. The optomechanical
coupling strengths are denoted by $g_{j}$. In order to generate steady
state entanglement, we assume a strong coherent drive on each cavity,
detuned to the red (blue) mechanical sideband for cavity 1(2), i.e.,
drive frequency $\omega_{\mathbf{d}1}=\omega_{1}-\omega_{\mathbf{m}},$
$\omega_{\mathbf{d}2}=\omega_{2}+\omega_{\mathbf{m}}$. We work in
an interaction picture with respect to free Hamiltonian and split
the cavity field into a classical cavity amplitude $\bar{a}_{j}$ and a small
quantum fluctuation $\hat{d}_{j}$ with $\hat{d}_{j}=\hat{a}_{j}-\bar{a}_{j}$. Here $\bar{a}_{j}$ is the average number of photons for each cavity. After performing the rotating wave approximation and linearizing the Hamiltonian
independently, We obtain:
\begin{equation}
\hat{H}_{\mathbf{int}}=\left(G_{1}\hat{b}^{\dagger}\hat{d}_{1}+G_{2}\hat{b}\hat{d}_{2}\right)+\mathbf{h.c.}
\end{equation}
Here $G_{j}=g_{j}\bar{a}_{j}$ is the dressed coupling. In general,
we take $g_{j},\bar{a}_{j}>0$. Taking the damping and the noise terns
into account, we get the Langevin equation for the optical and mechanical
modes operator \cite{CWG-QN-2004}:
\begin{align}
&\frac{d}{dt}\hat{b}  =-\frac{\gamma}{2}\hat{b}-\mathbf{i}(G_{1}\hat{d}_{1}+G_{2}\hat{d}_{2}^{\dagger})-\sqrt{\gamma}\hat{b}^{\mathbf{in}}(t), \nonumber\\
&\frac{d}{dt}\hat{d}_{1}  =-\frac{\kappa_{1}}{2}\hat{d}_{1}-\mathbf{i}G_{1}\hat{b}-\sqrt{\kappa_{1}}\hat{d}_{1}^{\mathbf{in}}(t), \nonumber\\
&\frac{d}{dt}\hat{d}_{2}^{\dagger}  =-\frac{\kappa_{2}}{2}\hat{d}_{2}^{\dagger}+\mathbf{i}G_{2}\hat{b}-\sqrt{\kappa_{2}}\hat{d}_{2}^{\mathbf{in},\dagger}(t),
\label{Langevin}
\end{align}
where $\kappa_{j}$ and $\gamma$ is the damping rate of the cavities
and mechanics respectively and $\mathbf{i}$ is the imaginary unit. As discussed in \cite{CWG-QN-2004}, from
the Langevin equations and the input-output relation, one can verify
that the stationary output state in the Fock state basis $\left|n_{1},n_{2},n_{\mathbf{m}}\right\rangle $,
can be expressed as:
\begin{equation}
\left|\Psi\right\rangle =\sum_{p,q}\frac{\sqrt{C_{p+q}^{p}}N_\mathbf{m}^{\frac{q}{2}}N_\mathbf{1}^{\frac{p}{2}}}{\left(1+N_\mathbf{2}\right)^{\left(p+q+1\right)/2}}\left|p,p+q,q\right\rangle .
\end{equation}
In Eq. {(4)}, $C_{p+q}^{p}$ is the binomial coefficients and $N_\mathbf{k} (k=1,2,m)$
is the average photons or phonons number of each output mode with
$N_\mathbf{k}=\left\langle \left(\hat{d}_{k}^{\mathbf{out}}\left[0\right]\right)^{\dagger}\hat{d}_{k}^{\mathbf{out}}\left[0\right]\right\rangle $
and they can be given as follow: 
\begin{align}
N_\mathbf{1}=\frac{4C_{1}C_{2}}{\left(1+C_{1}-C_{2}\right)^{2}}, \\
N_\mathbf{2}=\frac{4C_{2}\left(C_{1}+1\right)}{\left(1+C_{1}-C_{2}\right)^{2}},\\
N_\mathbf{m}=\frac{4C_{2}}{\left(1+C_{1}-C_{2}\right)^{2}},
\end{align}
with the cooperativity $C_{j}=4G_{j}^{2}/\gamma\kappa_{j}$.
The result was derived under the assumptions of zero temperature
and in the limit of narrow bandwidth around the bare frequencies $\omega_{j}, \omega_\mathbf{m}$.
Note that this is a Gaussian state (more specifically), it is a twice
squeezed 3-mode vacuum state \cite{YDW-PRA-2015}, which is a genuine
tripartite entangled state. By tracing out the mechanical  mode, one
obtains a 2-mode squeezed thermal state of the photon output fields,
which has entanglement:
\begin{equation}
E_{N}=\ln\frac{\left(1+C_{1}-C_{2}\right)^{2}}{A+B+2C_{2}\left(1+2C_{1}\right)-4\sqrt{AB}},
\end{equation}

with $A=C_{2}\left(C_{1}+C_{2}\right), B=\left(1+C_{1}\right)^{2}+C_{1}C_{2}$. The entanglement
is maximized at the instability point $C_{1}\rightarrow C_{2}-1$
and remains finite ($E_{N}\rightarrow\ln\left(2C_{1}+1\right)$),
at variance with the well-known divergence for a parametric amplifier.
This is a natural result because the two modes are entangled with
the mechanics. Indeed, the divergence is manifested only for the tripartite
entanglement \cite{YDW-PRA-2015}. However, as we discuss in the following,
a divergence of $E_{N}$ can be recovered by an ideal measurement.
In this sense, the large entanglement of the three-body state is a
physical resource which can be used to greatly enhance the bipartite
entanglement of the emitted phonons.

In the practice, the mechanics can be connected to a strong damped auxiliary
cavity (cf. Fig. 1) such that the mechanical output can be mapped
to the optical output, $\hat{d}_{a,\mathbf{out}}=-\mathbf{i}\hat{b}_\mathbf{out}$ \cite{YDW-PRA-2015}.
A recent experiment has demonstrated the readout of the phonon number
through this mechanism \cite{JDC-Nature-2015}. So in the following text, the measurement of phonon of the mechanical mode is through the measurement of photon of the auxiliary cavity mode. We will simply refer to this method as \textquotedbl{}measurement
of the phonon mode\textquotedbl{} and quantify its effect on the output
entanglement of the two cavities.

In the measurement theory of quantum mechanics, projection operator
is a perfect measurement operator
\begin{equation}
\hat{M}_{1}\left(q\right)=\left|q\right\rangle \left\langle q\right|,
\end{equation}
but in experiment we often deal with imperfect measurement. A typical
imperfect measurement is efficient measurement that the detect efficiency
$\mu$ is considered \cite{CAH-PRA-1989}. For a single photon detector,
the detect efficiency $\mu$ can be regarded as the probability for
detecting one photon in time t from an one photon field. The explicit
form of $\mu$ depends on the physical situation, here we just consider
a constant value of $\mu$. According to ref. \cite{CAH-PRA-1989}, the
operator of measurement with measure outcome $q$ is 
\begin{equation}
\hat{M}_{\mu}\left(q\right)=\sum_{n=q}^{\infty}\sqrt{C_{n}^{q}}\left(1-\mu\right)^{\frac{n-q}{2}}\mu^{\frac{q}{2}}\left|n-q\right\rangle \left\langle q\right|.
\end{equation}
We find that the imperfect measurement become perfect measurement
when $\mu=1$. Another measurement may be on-off measurement. The
on case can be interpreted as: we detect photon, but we can't identify
the photon number. The off case is that we have not detected photon.
In physics, they can be expressed as \cite{SLZ-PRA-2010}:
\begin{align}
&\hat{M}_{\mathbf{off}}=\left|0\right\rangle \left\langle 0\right|,\\
&\hat{M}_{\mathbf{on}}=\mathbf{I}-\left|0\right\rangle \left\langle 0\right|=\sum_{k=1}^{\infty}\left|k\right\rangle \left\langle k\right|.
\label{Langevin}
\end{align}

\section{perfect measurement}

We first consider the perfect measurement of the phonon number, described
by projection operator $\hat{M_{1}}(q)$ where $q$ is the outcome
of the phonon measurement. Such measurement increases the entanglement
\cite{WM-SCPMA-2015}, as it can be computed straightforwardly from
the state after measurement:
\begin{equation}
\left|\Psi_{q}\right\rangle =P_{q}^{-\frac{1}{2}}\hat{M_{1}}\left(q\right)\left|\Psi\right\rangle =\sum_{p=0}^{\infty}\sqrt{f_{p}\left(q\right)}\left|p,p+q\right\rangle .
\end{equation}
Here we define 
\begin{equation}
f_{p}\left(q\right)=C_{p+q}^{p}\zeta^{p}\left(1-\zeta\right)^{1+q},
\end{equation}
where $\zeta=4C_{1}C_{2}(1+C_{1}+C_{2})^{-2}$, The normalization
factor is 
\begin{equation}
P_{q}=\left\langle \Psi\right|\hat{M}_{1}^{\dagger}\left(q\right)\hat{M}_{1}\left(q\right)\left|\Psi\right\rangle =\frac{N_\mathbf{m}^{q}}{\left(1+N_\mathbf{m}\right)^{1+q}}.
\end{equation}
Although Eq. {(13)} is not a Gaussian state, we can still quantify
the entanglement directly from the definition of logarithmic negativity
\cite{GV-PRA-2002}:
\begin{equation}
E_{N}\equiv\ln\left\Vert \hat{\rho}^{\mathbf{T}}\right\Vert _{\mathbf{1}},
\end{equation}
where $\hat{\rho}$ is the density matrix of the state being evaluated,
$\hat{\rho}^{\mathbf{T}}$ is the partial transpose with respect to one
subsystem, $\left\Vert .\right\Vert _{\mathbf{1}}$ denotes trace
norm.  Also noticing that for a two mode entangled state written in a
Schmidt decompostion as $\left|\varphi\right\rangle =\sum C_{n}\left|n_{\mathbf{A}},n_{\mathbf{B}}\right\rangle $,
the entanglement is: $E_{N}=2\ln\sum_{n}\left|C_{n}\right|$. Thus the
entanglement of the state in Eq. {(13)} can be written as:
\begin{equation}
E_{N}\left(q\right)=2\ln\sum_{p=0}^{\infty}\sqrt{f_{p}\left(q\right)}.
\end{equation}
A special case is $q=0$, it means that no phonon has been detected.
In this case,
\begin{equation}
E_{N}\left(0\right)=\ln\frac{1+\sqrt{\zeta}}{1-\sqrt{\zeta}}.
\end{equation}
To get an analytical expression of the entanglement after measure, an approximation
is necessary. Notice that $f_{p}(q)$ is normalized and it can be
regarded as a Gaussian distribution for large $q$:
\begin{equation}
f_{p}\left(q\right)\approx\frac{1}{\sqrt{2\pi}\sigma\left(q\right)}e^{-\frac{\left[p-\kappa\left(q\right)\right]^{2}}{2\sigma\left(q\right)^{2}}}
\end{equation}
with the mean and variance being $\kappa\left(q\right)=\zeta\frac{1+q}{1-\zeta}$,
$\sigma\left(q\right)=\frac{\sqrt{\zeta\left(1+q\right)}}{1-\zeta}$.
Then:
\begin{equation}
E_{N}\left(q\right)\approx\ln\frac{\sqrt{8\pi\zeta\left(1+q\right)}}{1-\zeta}.
\end{equation}
     The entanglement after measurement is found to increase logarithmically
with the number of detected phonons ($q$), and it is larger than
the entanglement before measurement, even the measurement outcome is zero
(see Fig. 2 ). It can been found that measurement enhance entanglement.
This is one of the most important conclusion in the perfect
measurement.

\begin{figure}
\includegraphics[width=8cm]{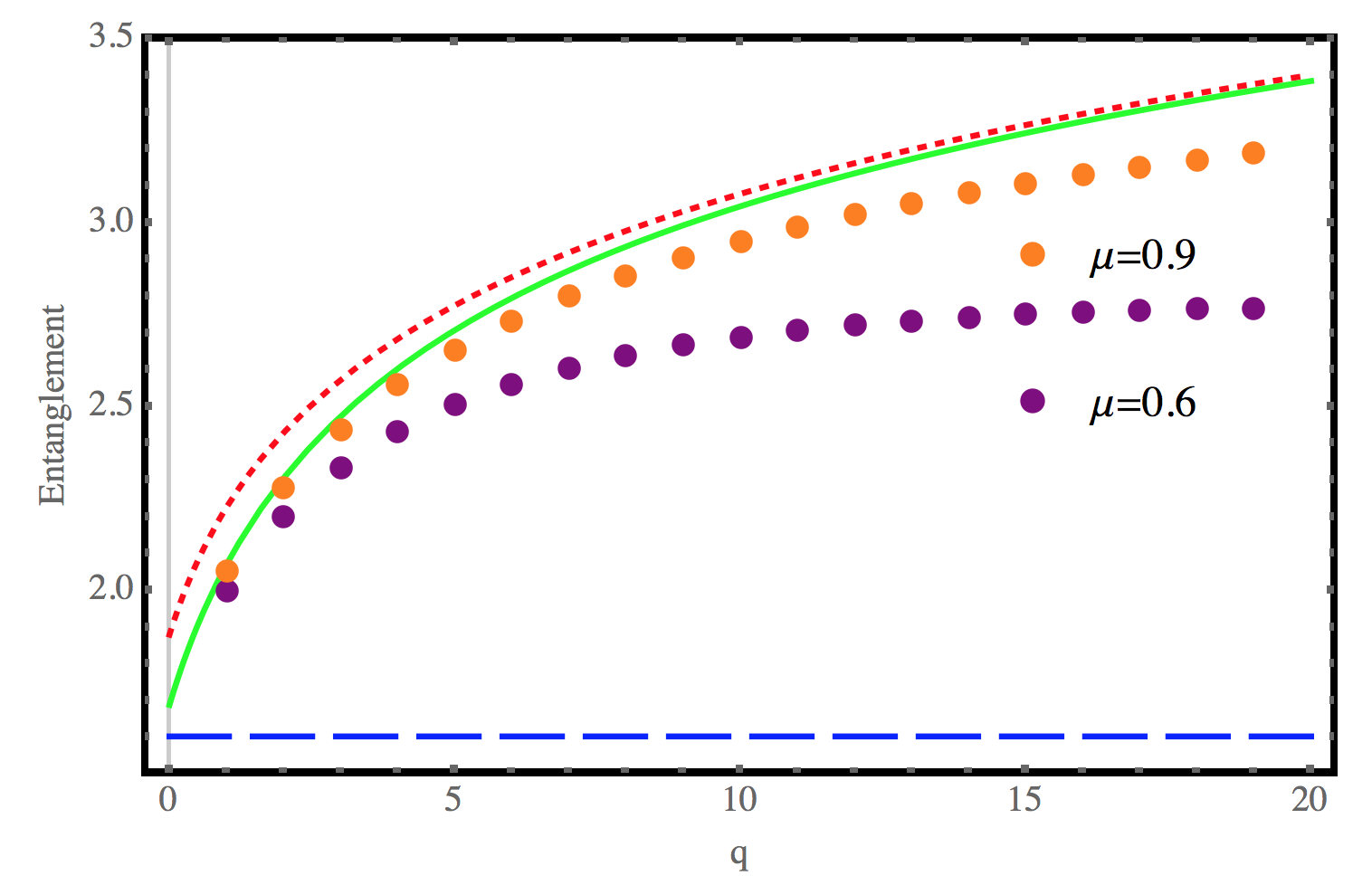}\caption{(Color online) Entanglement versus detected phonons$(q)$. Entanglement of the three mode state before measurement $E_{N}$(blue dash line),
entanglement after the perfect measurement $E_{N}\left(q\right)$(green line), entanglement with Gaussian approximation after the perfect measurement(red dot line), entanglement after the imperfect measurement with
$\mu=0.6$(purple dots) and entanglement after the imperfect measurement
with $\mu=0.9$(orange dots). It can be seen even the measurement
outcome is zero, $E_{N}\left(q\right)$ is larger than $E_{N}$. The
entanglement is concentrated. The parameters values is $C_{1}=10,C_{2}=2$.}
\end{figure}

\section{imperfect measurement}

Now we consider the imperfect measurement with the measurement operator
$\hat{M}_{\mu}\left(q\right)$. According to ref. \cite{CAH-PRA-1989},
the state after measurement is given by 
\begin{equation}
\hat{\rho}_{q}=\frac{1}{P_{\mu}\left(q\right)}Tr_{q}\left[\hat{M}_{\mu}\left(q\right)\hat{\rho}\hat{M}_{\mu}^{\dagger}\left(q\right)\right],
\end{equation}
here the trace is for detected mode, and $P_{\mu}(q)$ is the probability
for detecting $q$ phonons from a field with the phonon number distribution
$P_{s}$:
\begin{equation}
P_{\mu}\left(q\right)=\sum_{s=q}^{\infty}P_{s}C_{s}^{q}\mu^{q}\left(1-\mu\right)^{s-q}.
\end{equation}
For the given three mode entangled state $\left|\Psi\right\rangle $,
when we consider the detect efficiency $\mu$,
the state after the imperfect measurement with the detect outcome $q$
is:
\begin{equation}
\begin{split}\hat{\rho}_{q} & =\sum_{s=q}^{\infty}\eta\left(s\right)\left|\Psi_{s}\right\rangle \left\langle \Psi_{s}\right|\\
 & =\sum_{p_{1}=0}^{\infty}\sum_{,p_{2}=0}^{\infty}\sum_{s=q}^{\infty}\sqrt{f_{p_{1}}\left(s\right)f_{p_{2}}\left(s\right)}\eta\left(s\right)\\
 & \times\left|p_{1},p_{1}+s\right\rangle \left\langle p_{2},p_{2}+s\right|,
\end{split}
\end{equation}
with 
\begin{equation}
\eta\left(s\right)=\frac{N_\mathbf{m}^{s-q}\left(1+N_\mathbf{m}\mu\right)^{1+q}}{\left(1+N_\mathbf{m}\right)^{1+s}}C_{s}^{q}\left(1-\mu\right)^{s-q}.
\end{equation}
Obvious, $\hat{\rho}_{q}$ is a mixed state. From the ref.  \cite{GV-PRA-2002},
calculating logarithmic entanglement is to find the negativity $N\left(\hat{\rho}_{q}^{\mathbf{T}}\right)$
with 
\begin{equation}
\begin{split}\hat{\rho}_{q}^{\mathbf{T}} & =\sum_{p_{1}=0}^{\infty}\sum_{,p_{2}=0}^{\infty}\sum_{s=q}^{\infty}\sqrt{f_{p_{1}}\left(s\right)f_{p_{2}}\left(s\right)}\eta\left(s\right)\\
 & \times\left|p_{1},p_{2}+s\right\rangle \left\langle p_{2},p_{1}+s\right|.
\end{split}
\end{equation}
The $\hat{\rho}_{q}^{\mathbf{T}}$ is a block diagonal matrix
by identifying the different total photons $Q=p_{1}+p_{2}+s$ for each subblock. When
$Q=q$, there is only one matrix element $Q\left[q\right]=f_{0}(q)\eta\left(q\right)$
in the subblock matrix. When $Q=q+1$, it is a $2\times2$ square
matrix with $Q\left[q+1\right]$:
\begin{equation}
Q\left[q+1\right]=\left[\begin{array}{cc}
\sqrt{f_{1}\left(q\right)f_{0}\left(q\right)}\eta\left(q\right) & 0\\
f_{0}\left(q+1\right)\eta\left(q+1\right) & \sqrt{f_{0}\left(q\right)f_{1}\left(q\right)}\eta\left(q\right)
\end{array}\right].
\end{equation}
When $Q=q+n$, the sub block matrix $Q\left[q+n\right]$ is a $n\times n$
square matrix. Each sub block matrix of $\hat{\rho}_{q}^{\mathbf{T}}$ is a lower
triangular matrix. To calculate $N\left(\hat{\rho}_{q}^{\mathbf{T}}\right)$,
we make an approximation. By throwing away most elements and only remaining
the main diagonal and the elements which are the nearest to the main
diagonal for each sub blocks,  then the expression of  $\hat{\rho}_{q}^{\mathbf{T}}$ becomes
\begin{equation}
\begin{split}\hat{\rho}_{q}^{\mathbf{T}} & =\sum_{p_{1}=0}^{\infty}\sum_{,p_{2}=0}^{\infty}\sum_{s=q}^{q+1}\sqrt{f_{p_{1}}\left(s\right)f_{p_{2}}\left(s\right)}\eta\left(s\right)\\
 & \times\left|p_{1},p_{2}+s\right\rangle \left\langle p_{2},p_{1}+s\right|.
\end{split}
\end{equation}
Then calculating the entanglement after the imperfect measurement is calculating the eigenvalues of
$\hat{\rho}_{q}^{\mathbf{T}}$.  To calculate the eigenvalues of
$\hat{\rho}_{q}^{\mathbf{T}}$, we thought about using perturbation theory.
In quantum mechanics, the classical non degenerate stationary state
perturbation theory is that for a given $\hat{H}$, we can devide
$\hat{H}$ into two part: $\hat{H}=\hat{H}_{\mathbf{0}}+\hat{H}^{\mathbf{'}}$($\hat{H}^{\mathbf{'}}$
is a perturbation). Here, we just treat $\hat{\rho}_{q}^{\mathbf{T}}$
as $\hat{H}$ and let $\hat{\rho}_{q}^{\mathbf{T}}=\hat{H}=\hat{H}_{\mathbf{0}}+\hat{H}^{\mathbf{'}}$
with
\begin{equation}
\begin{split}\hat{H}_{\mathbf{0}} & =\sum_{p_{1}=0}^{\infty}\sum_{,p_{2}=0}^{\infty}\sqrt{f_{p_{1}}\left(q\right)f_{p_{2}}\left(q\right)}\eta\left(q\right)\\
 & \times\left|p_{1},p_{2}+q\right\rangle \left\langle p_{2},p_{1}+q\right|,
\end{split}
\end{equation}
\begin{equation}
\begin{split}\hat{H}^{\mathbf{'}} & =\sum_{p_{1}=0}^{\infty}\sum_{,p_{2}=0}^{\infty}\sqrt{f_{p_{1}}\left(q+1\right)f_{p_{2}}\left(q+1\right)}\eta\left(q+1\right)\\
 & \times\left|p_{1},p_{2}+q+1\right\rangle \left\langle p_{2},p_{1}+q+1\right|.
\end{split}
\end{equation}
Also for a classical perturbation theory, the eigen equation of $\hat{H}_{\mathbf{0}}$
is $\hat{H}_{\mathbf{0}}\left|m\right\rangle =e_{m}^{(0)}\left|m\right\rangle $, where $e_{m}^{(0)}$ is eigenvalue in the eigenstate $\left|m\right\rangle $.
The first-order approximate eigenvalue of $\hat{H}$ in a state $\left|\phi_{m}\right\rangle $
(which is close to $\left|m\right\rangle $):
\begin{equation}
e_{m}^{(1)}=\left\langle m\right|\hat{H}^{\mathbf{'}}\left|m\right\rangle ,
\end{equation}
and the second-order approximate eigenvalue of $\hat{H}$ in a state
$\left|\phi_{m}\right\rangle $:
\begin{equation}
e_{m}^{(2)}=\sum_{k\neq m}\frac{\left[\left\langle m\right|\hat{H}^{\mathbf{'}}\left|k\right\rangle \right]^{2}}{e_{m}^{\left(0\right)}-e_{k}^{\left(0\right)}}.
\end{equation}
Then we have the approximate eigenvalue of $\hat{H}$ in a state $\left|\phi_{m}\right\rangle $
\begin{equation}
e_{m}=e_{m}^{(0)}+e_{m}^{(1)}+e_{m}^{(2)}.
\end{equation}
For $\hat{\rho}_{q}^{\mathbf{T}}$, its eigenvectors and eigenvalues
of $\hat{H}_{\mathbf{0}}$ is 
\begin{equation}
\begin{array}{ccc}
condition & eigenvectors & eigenvalues\\
p_{1}=p_{2}=p & \left|\alpha\right\rangle  & f_{p}\left(q\right)\eta\left(q\right)\\
p_{1}<p_{2} & \left|\beta_{+}\right\rangle  & +\sqrt{f_{p_{1}}\left(q\right)f_{p_{2}}\left(q\right)}\eta\left(q\right)\\
p_{1}<p_{2} & \left|\beta_{-}\right\rangle  & -\sqrt{f_{p_{1}}\left(q\right)f_{p_{2}}\left(q\right)}\eta\left(q\right)
\end{array}
\end{equation}
with $\left|\alpha\right\rangle =\mid p,p+q\rangle$, $\left|\beta_{\pm}\right\rangle =\frac{\mid p_{1},p_{2}+q\rangle\pm\mid p_{2},p_{1}+q\rangle}{\sqrt{2}}$.
According to the definition of $N\left(\hat{\rho}_{q}^{\mathbf{T}}\right)$,
only the eigenvalue of the state $\left|\beta_{-}\right\rangle $
is important. Comparing to the classic formula of perturbation theory,
the entanglement after imperfect measurement is:
\begin{equation}
E_{N}^{\mu}\left({q}\right)=\ln\left[1+2\sum_{\beta_{-}}\left|E_{\beta_{-}}\right|\right],
\end{equation}
with
\begin{equation}
\begin{split}E_{\beta_{-}}= & E_{\beta_{-}}^{\left(0\right)}+E_{\beta_{-}}^{\left(1\right)}+E_{\beta_{-}}^{\left(2\right)}\\
E_{\beta_{-}}^{\left(0\right)}= & -\sqrt{f_{p_{1}}\left(q\right)f_{p_{2}}\left(q\right)}\eta\left(q\right)\\
E_{\beta_{-}}^{\left(1\right)}= & \left\langle \beta_{-}\right|\hat{H}^{\mathbf{'}}\left|\beta_{-}\right\rangle \\
E_{\beta_{-}}^{\left(2\right)}= & \sum_{m\neq\beta_{-}}\frac{\left[\left\langle \beta_{-}\right|\hat{H}^{\mathbf{'}}\left|m\right\rangle \right]^{2}}{E_{\beta_{-}}^{\left(0\right)}-E_{m}^{\left(0\right)}}.
\end{split}
\end{equation}
When only consider the first order, the entanglement can be expressed
as:
\begin{equation}
E_{N}^{\mu}\left({q}\right)=E_{N}\left(q\right)-\left(q+1\right)\varepsilon,
\end{equation}

with $\varepsilon=\frac{\left(1-\mu\right)N_\mathbf{m}}{\left(1+N_\mathbf{m}\right)}$. Considering
the second order, it is:
\begin{equation}
E_{N}^{\mu}\left({q}\right)=E_{N}\left(q\right)+\left(q+1\right)\varepsilon\left[\frac{q\varOmega+\varOmega-1}{2}\varepsilon-1\right],
\end{equation}
with 
\begin{equation}
\varOmega=\frac{\sum_{p_{1}=0}^{\infty}\sum_{p_{2}=0}^{\infty}g\left(p_{1},p_{2}\right)}{\left(\sum_{p=0}^{\infty}\sqrt{f_{p}\left(q\right)}\right)^{2}},
\end{equation}
\begin{equation}
g\left(p_{1},p_{2}\right)=\frac{f_{p_{2}}\left(q+1\right)f_{p_{1}}\left(q+1\right)}{\sqrt{f_{p_{1}}\left(q\right)f_{p_{2}+1}\left(q\right)}+\sqrt{f_{p_{1}+1}\left(q\right)f_{p_{2}}\left(q\right)}}.
\end{equation}
In the Gaussian approximation, $\Omega$ can get a simple expression,
and when $q$ is large, $\Omega$ will tend to $\frac{1}{2}$,
\begin{equation} 
\varOmega\approx\frac{1}{2}\sqrt{\frac{1+\zeta q}{\zeta+\zeta q}}\approx\frac{1}{2}. 
\end{equation}
Then we get 
\begin{equation}
E_{N}^{\mu}\left({q}\right)=E_{N}\left(q\right)-\left(q+1\right)\varepsilon+\frac{q^{2}-1}{4}\varepsilon^{2}.
\end{equation}

In the numerical analysis of the entanglement after the imperfect measurement (see Fig. 2), 
we can see that the imperfect measurement will be close to the perfect measurement when $\mu$ is close to 1.
 In the analytical analysis, the perturbation approximation is effective when $\mu\rightarrow1$ and the detected phonon number $q$ is small (see Fig. 4).
It can be found that the entanglement after imperfect measurement is
larger than the entanglement before measurement, but smaller than the
entanglement after perfect measurement (see Fig. 3). That is to say:
perfect measurement is more effective than imperfect measurement in the
entanglement concentration, and no matter adopt what kind of measurement
method, the entanglement is concentred.

\begin{figure}
\includegraphics[width=8cm]{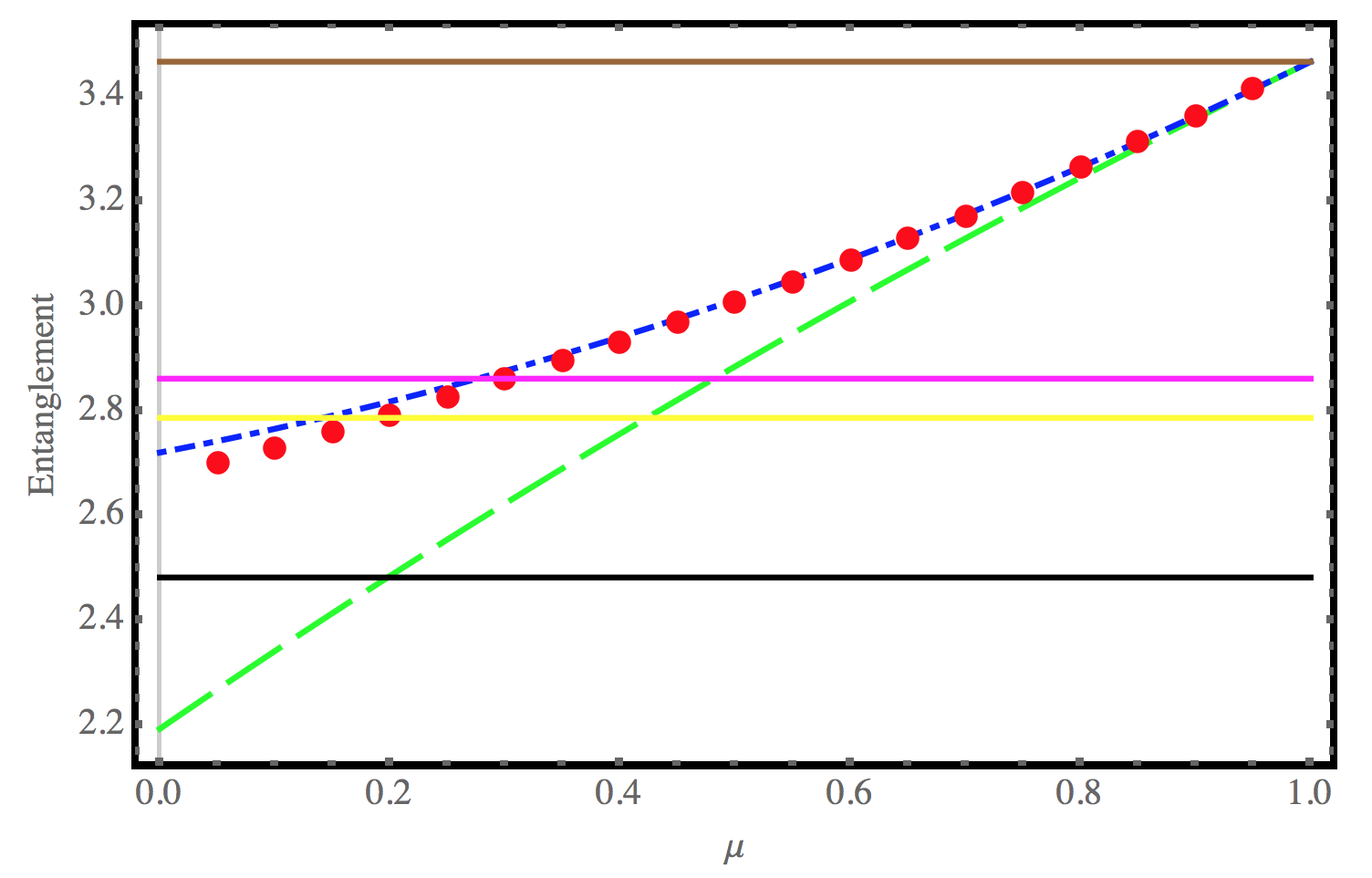}\caption{(Color online) Entanglement versus detect efficiency $\mu$. Entanglement
after the perfect measurement $E_{N}\left(q\right)$(brown line), the
three mode state entanglement before measurement $E_{N}$(black line),
entanglement after on measurement $E_{N}^{\mathbf{on}}$(pink line),
entanglement after off measurement $E_{N}^{\mathbf{off}}$(yellow line),
entanglement with numerical results after the imperfect measurement(red
dot), the first order approximation of the entanglement after imperfect
measurement(green dash line) and the second order approximation
of the entanglement after imperfect measurement(blue dot line). It can be seen the perturbation
approximation is effective and the imperfect measurement will be close
to the perfect measurement when $\mu\rightarrow1$. The
parameters values is $C_{1}=10,C_{2}=5,q=2$.}
\end{figure}

\begin{figure}
\includegraphics[width=16cm]{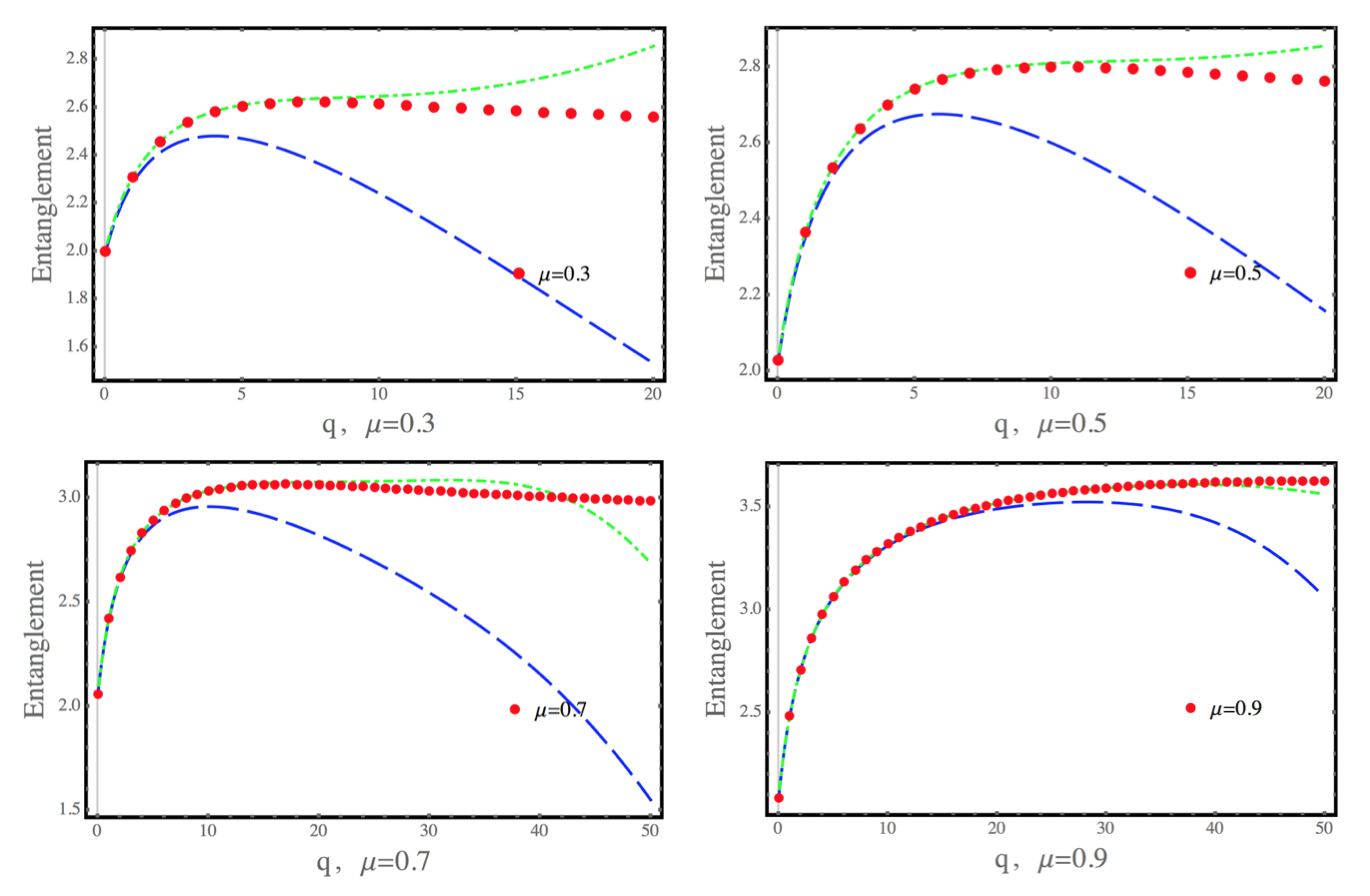}\caption{(Color online) Entanglement versus detected phonons (q) with different detect efficiency $\mu$. 
Entanglement with numerical results after the imperfect measurement(red
dot), the first order approximation of the entanglement after imperfect
measurement(blue dash line) and the second order approximation
of the entanglement after imperfect measurement(green dash line)
The parameters values is $C_{1}=10,C_{2}=3$.}
\end{figure}

\section{on-off measurement}

\begin{figure}
\includegraphics[width=8cm]{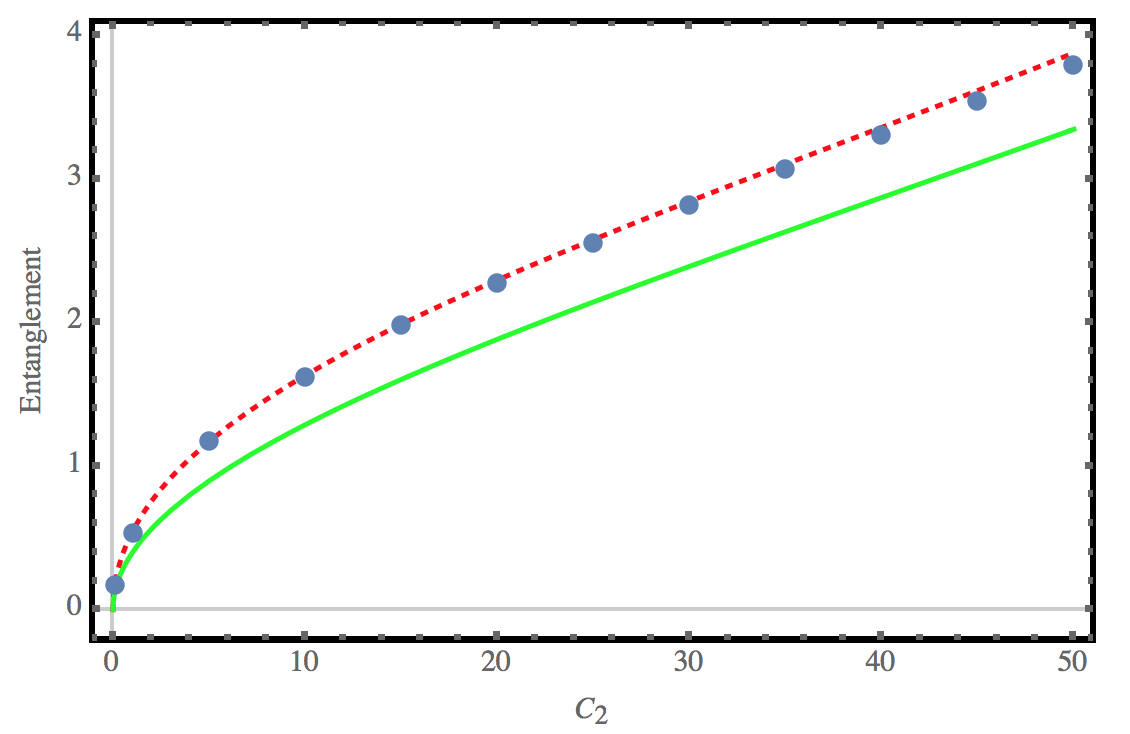}\caption{(Color online) Entanglement versus cooperativities $C_{2}$. The
numerical result for on measurement entanglement $E_{N}^{\mathbf{on}}$
(blue dots), the average entanglement$\overline{E}_{N}^{\mathbf{on}}$
(red dash line), the three mode state entanglement before measurement $E_{N}$
(green line). The parameters values is $C_{1}=100$.}
\end{figure}

Now we consider on-off measurement and the state after measure is 
\begin{align}
&\hat{\rho}_{\mathbf{off}}=\left|\Psi_{0}\right\rangle \left\langle \Psi_{0}\right|,\\
&\hat{\rho}_{\mathbf{on}}=\sum_{k=1}^{\infty}\frac{1+N_\mathbf{m}}{N_\mathbf{m}}P_{k}\left|\Psi_{k}\right\rangle \left\langle \Psi_{k}\right|.
\end{align}
For off case, the entanglement is the same with the perfect measurement
when $q=0$. 
\begin{equation}
E_{N}^{\mathbf{off}}=E_{N}\left(0\right)=\ln\frac{1+\sqrt{\zeta}}{1-\sqrt{\zeta}}.
\end{equation}
For on case, the logarithmic negativity theory still works. We can
calculate the $N\left(\hat{\rho}_{\mathbf{on}}^{\mathbf{T}}\right)$. Then we get $E_{N}^{\mathbf{on}}=\ln\left(1+2N\left(\hat{\rho}_{\mathbf{on}}^{\mathbf{T}}\right)\right)$.
In this way it is easy to get a precise numerical solution but difficult
to find an analytical expression. Fortunately we find that when make
the suitable parameters($C_{1}\gg C_{2}$), the average entanglement
is close to the numerical on measurement entanglement (see Fig. 5)
\begin{equation}
E_{N}^{\mathbf{on}}\approx\overline{E}_{N}^{\mathbf{on}},
\end{equation}
and the average entanglement is defined as:
\begin{equation}
\overline{E}_{N}^{\mathbf{on}}=\frac{1+N_\mathbf{m}}{N_\mathbf{m}}\sum_{k=1}^{\infty}P_{k}E_{N}\left(k\right).
\end{equation}
If use Gaussian approximation:
\begin{equation}
\overline{E}_{N}^{\mathbf{on}}\approx\sum_{k=1}^{\infty}\frac{1}{N_\mathbf{m}}\left(\frac{N_\mathbf{m}}{1+N_\mathbf{m}}\right)^{k}\ln\frac{\sqrt{8\pi\zeta\left(1+k\right)}}{1-\zeta}.
\end{equation}
We find even in the on-off measurement, the entanglement is concentred,
but the effect is weakest comparing to the perfect measurement and the imperfect
measurement (see Fig. 3).

\section{conclusion}

In quantum communication, we often need the maximum entangled state. How to achieve maximum entangled state is an important subject of quantum communication. Entanglement concentration is an important method in preparing the maximum entangled state. In the continuous-variable, for the Gaussian entangled state, only the non Gaussian operation is possible in the entanglement concentration. As a non Gaussian operation-quantum measurement, we use three different measurement operators(perfect measurement, imperfect measurement and on-off measurement) to concentrate a 3-mode Gaussian state. Perfect measurement is strongest in the entanglement concentration, imperfect measurement is second, and the on-off measurement is the weakest. But on the other hand, perfect measurement get the most precise measurement information. In this respect, the more precise measurement information, the larger entanglement concentration. But In the experiment the imperfect measurement is relatively easy to implementation, so making an efficient measuring instrument is very important. In the imperfect measurement, we use perturbation theory to calculate the entanglement of a mixed state, the numerical results and analytical results fit well. This has some reference for us to deal with the entanglement of other density matrices. No matter the perfect measurement, imperfect measurement and on-off measurement, they are strong measurement, the entanglement concentration based on weak measurement will be considered in the next work.

\newpage

\section{SUPPLEMENT}

\subsection*{Supplement[A]: The 3-mode Entangled State}

The 3-mode entangled state is 
\begin{equation}
\left|\Psi\right\rangle =\sum_{p,q=0}^{\infty}\sqrt{\frac{C_{p+q}^{p}}{1+N_{2}}}\left(\frac{N_{m}}{1+N_{2}}\right)^{\frac{q}{2}}\left(\frac{N_{1}}{1+N_{2}}\right)^{\frac{p}{2}}\left|p,p+q,q\right\rangle ,
\end{equation}
with 
\begin{equation}
\begin{array}{c}
N_{1}=\frac{4C_{1}C_{2}}{\left(1+C_{1}-C_{2}\right)^{2}},\\
N_{2}=\frac{4C_{2}\left(C_{1}+1\right)}{\left(1+C_{1}-C_{2}\right)^{2}},\\
N_{m}=\frac{4C_{2}}{\left(1+C_{1}-C_{2}\right)^{2}}.
\end{array}
\end{equation}
Obvious we have $N_{2}=N_{1}+N_{m}$. For the 3-mode entangled state,
$p$ is one cavity mode, $p+q$ is another cavity mode, $q$ is mechanical
mode. So $p,p+q$ represent the photon, $q$ represent the phonon.
In order to make the following calculation process simple, we let
\begin{equation}
P_{q}f_{p}\left(q\right)=\frac{C_{p+q}^{p}}{1+N_{2}}\left(\frac{N_{m}}{1+N_{2}}\right)^{q}\left(\frac{N_{1}}{1+N_{2}}\right)^{p}.
\end{equation}
The definition of $f_{p}\left(q\right)$ and $P_{q}$ are described
in the following part. Then we can make a short expression for $\left|\Psi\right\rangle $:
\begin{equation}
\left|\Psi\right\rangle =\sum_{p,q=0}^{\infty}\sqrt{P_{q}f_{p}\left(q\right)}\left|p,p+q,q\right\rangle 
\end{equation}
and the density matrix is 
\begin{equation}
\begin{split}\rho_{i}=\left|\Psi\right\rangle \left\langle \Psi\right|= & \sum_{p_{1},q_{1}=0}^{\infty}\sum_{p_{2},q_{2}=0}^{\infty}\frac{\sqrt{C_{p_{1}+q_{1}}^{p_{1}}C_{p_{2}+q_{2}}^{p_{2}}}}{1+N_{2}}\left(\frac{N_{m}}{1+N_{2}}\right)^{\frac{q_{1}+q_{2}}{2}}\left(\frac{N_{1}}{1+N_{2}}\right)^{\frac{p_{1}+p_{2}}{2}}\\
 & \times\left|p_{1},p_{1}+q_{1},q_{1}\right\rangle \left\langle p_{2},p_{2}+q_{2},q_{2}\right|
\end{split}
\end{equation}

also
\begin{equation}
\begin{split}\rho_{i}= & \sum_{p_{1},q_{1}=0}^{\infty}\sum_{p_{2},q_{2}=0}^{\infty}\sqrt{P_{q_{1}}f_{p_{1}}\left(q_{1}\right)}\sqrt{P_{q_{2}}f_{p_{2}}\left(q_{2}\right)}\\
 & \times\left|p_{1},p_{1}+q_{1},q_{1}\right\rangle \left\langle p_{2},p_{2}+q_{2},q_{2}\right|
\end{split}
\end{equation}
and the entanglement of the two cavity mode is 
\begin{equation}
E_{N}=\ln\frac{\left(1+C_{1}-C_{2}\right)^{2}}{A+B+2C_{2}\left(1+2C_{1}\right)-4\sqrt{AB}}
\end{equation}
with $A=C_{2}\left(C_{1}+C_{2}\right)$ , $B=\left(1+C_{1}\right)^{2}+C_{1}C_{2}$.

\subsection*{Supplement[B]: Phonon Measure with Perfect Measument}

If we measure the mechanical mode and measure $q$ phonons , the measurement
opreater is 

\begin{equation}
\hat{M_{1}}\left(q\right)=\left|q\right\rangle _{q}\left\langle q\right|
\end{equation}
and the quantum state after measure is given by 

\begin{equation}
\rho_{q}^{\prime}=Tr_{q}\left[\hat{M_{1}}\left(q\right)\rho_{i}\right]
\end{equation}
and the normalized state after measure is given by 
\begin{equation}
\rho_{q}=\frac{\rho_{q}^{\prime}}{Tr\left[\rho_{q}^{\prime}\right]}
\end{equation}
and the trace is calculated by the following
\begin{equation}
Tr_{q}\left[\hat{M_{1}}\left(q\right)\rho_{i}\right]=\sum_{q_{3}=0}^{\infty}\left\langle q_{3}\right|\hat{M_{1}}\left(q\right)\rho_{i}\left|q_{3}\right\rangle 
\end{equation}
After calculation we have 
\begin{equation}
\begin{split}\rho_{q}^{\prime}= & \sum_{p_{1}=0}^{\infty}\sum_{p_{2}=0}^{\infty}\frac{\sqrt{C_{p_{1}+q}^{p_{1}}C_{p_{2}+q}^{p_{2}}}}{1+N_{2}}\left(\frac{N_{m}}{1+N_{2}}\right)^{q}\left(\frac{N_{1}}{1+N_{2}}\right)^{\frac{p_{1}+p_{2}}{2}}\\
 & \times\left|p_{1},p_{1}+q\right\rangle \left\langle p_{2},p_{2}+q\right|
\end{split}
\end{equation}

we write as 
\begin{equation}
\rho_{q}^{\prime}=\left|\Psi_{q}^{\prime}\right\rangle \left\langle \Psi_{q}^{\prime}\right|
\end{equation}
with 
\begin{equation}
\left|\Psi_{q}^{\prime}\right\rangle =\sum_{p=0}^{\infty}\sqrt{\frac{C_{p+q}^{p}}{1+N_{2}}}\left(\frac{N_{m}}{1+N_{2}}\right)^{\frac{q}{2}}\left(\frac{N_{1}}{1+N_{2}}\right)^{\frac{p}{2}}\left|p,p+q\right\rangle 
\end{equation}
the $\left|\Psi_{q}^{\prime}\right\rangle $ is not a normalized state.
So we normalize the $\left|\Psi_{q}^{\prime}\right\rangle $
\[
\left\langle \Psi_{q}^{\prime}\mid\Psi_{q}^{\prime}\right\rangle =\sum_{p=0}^{\infty}\frac{C_{p+q}^{p}}{1+N_{2}}\left(\frac{N_{m}}{1+N_{2}}\right)^{q}\left(\frac{N_{1}}{1+N_{2}}\right)^{p}=P_{q}
\]
with
\begin{equation}
P_{q}=\frac{N_{m}^{q}}{\left(1-N_{1}+N_{2}\right)^{1+q}}=\frac{N_{m}^{q}}{\left(1+N_{m}\right)^{1+q}}
\end{equation}
and 
\begin{equation}
Tr\left[\rho_{q}^{\prime}\right]=P_{q}
\end{equation}
\begin{equation}
\rho_{q}=\frac{\rho_{q}^{\prime}}{Tr\left[\rho_{q}^{\prime}\right]}=\frac{\rho_{q}^{\prime}}{P_{q}}=\frac{1}{P_{q}}\left|\Psi_{q}^{\prime}\right\rangle \left\langle \Psi_{q}^{\prime}\right|
\end{equation}
so we can define 

\begin{equation}
\left|\Psi_{q}\right\rangle =\frac{1}{\sqrt{P_{q}}}\left|\Psi_{q}^{\prime}\right\rangle =\frac{1}{\sqrt{P_{q}}}\sum_{p=0}^{\infty}\sqrt{\frac{C_{p+q}^{p}}{1+N_{2}}}\left(\frac{N_{m}}{1+N_{2}}\right)^{\frac{q}{2}}\left(\frac{N_{1}}{1+N_{2}}\right)^{\frac{p}{2}}\left|p,p+q\right\rangle 
\end{equation}
with $\left|\Psi_{q}\right\rangle $ is a normalized state of $\left|\Psi_{q}^{\prime}\right\rangle $
\[
\left\langle \Psi_{q}\mid\Psi_{q}\right\rangle =1
\]
and
\begin{equation}
\rho_{q}=\left|\Psi_{q}\right\rangle \left\langle \Psi_{q}\right|
\end{equation}
so we have the normalized $\left|\Psi_{q}\right\rangle $ and it can
also be expressed as
\begin{equation}
\begin{split}\left|\Psi_{q}\right\rangle  & =\frac{1}{\sqrt{P_{q}}}\sum_{p=0}^{\infty}\sqrt{\frac{C_{p+q}^{p}}{1+N_{2}}}\left(\frac{N_{m}}{1+N_{2}}\right)^{\frac{q}{2}}\left(\frac{N_{1}}{1+N_{2}}\right)^{\frac{p}{2}}\left|p,p+q\right\rangle \\
 & =\sum_{p=0}^{\infty}\sqrt{C_{p+q}^{p}}\left(\frac{1+N_{m}}{1+N_{2}}\right)^{\frac{1+q}{2}}\left(\frac{N_{1}}{1+N_{2}}\right)^{\frac{p}{2}}\left|p,p+q\right\rangle \\
 & =\sum_{p=0}^{\infty}\sqrt{C_{p+q}^{p}}\left(1-\frac{N_{1}}{1+N_{2}}\right)^{\frac{1+q}{2}}\left(\frac{N_{1}}{1+N_{2}}\right)^{\frac{p}{2}}\left|p,p+q\right\rangle \\
 & =\sum_{p=0}^{\infty}\sqrt{C_{p+q}^{p}}\left(1-\zeta\right)^{\frac{1+q}{2}}\left(\zeta\right)^{\frac{p}{2}}\left|p,p+q\right\rangle \\
 & =\sum_{p=0}^{\infty}\sqrt{f_{p}\left(q\right)}\left|p,p+q\right\rangle 
\end{split}
\end{equation}
here we define
\begin{equation}
\zeta=\frac{N_{1}}{1+N_{2}}=\frac{4C_{1}C_{2}}{\left(1+C_{1}+C_{2}\right)^{2}}
\end{equation}
\begin{equation}
\begin{split}f_{p}\left(q\right) & =\frac{1}{P_{q}}\frac{C_{p+q}^{p}}{1+N_{2}}\left(\frac{N_{m}}{1+N_{2}}\right)^{q}\left(\frac{N_{1}}{1+N_{2}}\right)^{p}\\
 & =C_{p+q}^{p}\zeta^{p}\left(1-\zeta\right)^{1+q}
\end{split}
\end{equation}
 so we have the state after perfect measurement 
\begin{equation}
\left|\Psi_{q}\right\rangle =\sum_{p=0}^{\infty}\sqrt{f_{p}\left(q\right)}\left|p,p+q\right\rangle ,
\end{equation}
and the entanglement after perfect measurement is
\begin{equation}
E_{N}\left(q\right)=2\ln\sum_{p=0}^{\infty}\sqrt{f_{p}\left(q\right)}.
\end{equation}
To get an analytical expression of the entanglement after measure,
a approximation is necessary. Notice that $f_{p}\left(q\right)$ is
normalized and it can be regarded as a Gaussian distribution for large
$q$: 
\begin{equation}
f_{p}\left(q\right)\approx\frac{1}{\sqrt{2\pi}\sigma\left(q\right)}e^{-\frac{\left(p-\kappa\left(q\right)\right)^{2}}{2\sigma\left(q\right)^{2}}}.
\end{equation}
Comparing a standard Gaussian fuction 
\begin{equation}
G\left(x\right)=\frac{1}{\sqrt{2\pi}\sigma}e^{-\frac{(x-\mu)^{2}}{2\sigma^{2}}}
\end{equation}
we have 
\begin{equation}
\int G\left(x\right)dx=1
\end{equation}
and 
\begin{equation}
\left\langle x\right\rangle =\int G\left(x\right)xdx
\end{equation}
\begin{equation}
\left\langle x^{2}\right\rangle =\int G\left(x\right)x^{2}dx
\end{equation}
the mean and variance being is given by 
\begin{equation}
\mu=\left\langle x\right\rangle 
\end{equation}
\begin{equation}
\sigma=\sqrt{\left\langle x^{2}\right\rangle -\left\langle x\right\rangle ^{2}}
\end{equation}
So we have 
\begin{equation}
\sum_{p}f_{p}\left(q\right)=1
\end{equation}
\begin{equation}
\left\langle p\right\rangle =\sum_{p}f_{p}\left(q\right)p=\zeta\frac{1+q}{1-\zeta}
\end{equation}
\begin{equation}
\left\langle p^{2}\right\rangle =\sum_{p}f_{p}\left(q\right)p^{2}=\frac{\zeta\left(1+q\right)\left(1+\zeta+q\zeta\right)}{\left(1-\zeta\right)^{3}}
\end{equation}
and the mean and variance is given by 
\begin{equation}
\kappa\left(q\right)=\left\langle p\right\rangle =\zeta\frac{1+q}{1-\zeta}
\end{equation}
\begin{equation}
\sigma\left(q\right)=\sqrt{\left\langle p^{2}\right\rangle -\left\langle p\right\rangle ^{2}}=\frac{\sqrt{\zeta\left(1+q\right)}}{1-\zeta}
\end{equation}
Then after using Gaussian approximation, we have
\begin{equation}
E_{N}\left(q\right)\approx\ln\frac{\sqrt{8\pi\zeta\left(1+q\right)}}{1-\zeta}
\end{equation}

\subsection*{Supplement[C]: Phonon Measure with Imperfect Measument}

If the phonon detect effiency is $\mu$, and the state is in the $\left|\Psi_{s}\right\rangle $,
the probability for dectecting $q$ phonon is 

\begin{equation}
P\left(q,s\right)=C_{s}^{q}\mu^{q}\left(1-\mu\right)^{s-q}
\end{equation}
then calculate the $s$, get the probability for dectecting $q$ phonon
in $\left|\Psi\right\rangle $

\begin{equation}
P_{\mu}\left(q\right)=\sum_{s=q}^{\infty}P_{s}P\left(q,s\right)=\sum_{s=n}^{\infty}P_{s}C_{s}^{q}\mu^{q}\left(1-\mu\right)^{s-q}
\end{equation}
Simplify it we have 

\begin{equation}
P_{\mu}\left(q\right)=\frac{N_{m}^{q}\mu^{q}}{\left(1-N_{1}+N_{2}-N_{m}+N_{m}\mu\right)^{1+q}}=\frac{N_{m}^{q}\mu^{q}}{\left(1+N_{m}\mu\right)^{1+q}}
\end{equation}
When we have measure mechanical mode with the measurement outcome
$q$ , the condition of the two cavity mode after measure is given
by 

\begin{equation}
\rho_{q}=\frac{1}{P_{\mu}\left(q\right)}Tr_{q}\left[\hat{M_{\mu}}\left(q\right)\rho\hat{M_{\mu}^{\dagger}}\left(q\right)\right]
\end{equation}
where
\begin{equation}
\hat{M_{\mu}}\left(q\right)=\sum_{n=q}^{\infty}\sqrt{C_{n}^{q}}\left(1-\mu\right)^{\frac{n-q}{2}}\mu^{\frac{q}{2}}\left|n-q\right\rangle _{q}\left\langle n\right|
\end{equation}

\begin{equation}
\hat{M_{\mu}^{\dagger}}\left(q\right)=\sum_{n=q}^{\infty}\sqrt{C_{n}^{q}}\left(1-\mu\right)^{\frac{n-q}{2}}\mu^{\frac{q}{2}}\left|q\right\rangle _{q}\left\langle n-q\right|
\end{equation}
then the main task is to calculate $Tr_{q}\left[\hat{M_{\mu}}\left(q\right)\rho\hat{M_{\mu}^{\dagger}}\left(q\right)\right]$
and it can be written as

\begin{equation}
Tr_{q}\left[\hat{M_{\mu}}\left(q\right)\rho_{i}\hat{M_{\mu}^{\dagger}}\left(q\right)\right]=Tr_{q}\left[\hat{M_{\mu}}\left(q\right)\left|\Psi\right\rangle \left\langle \Psi\right|\hat{M_{\mu}^{\dagger}}\left(q\right)\right]
\end{equation}
and because $\left|\Psi\right\rangle $

\begin{equation}
\left|\Psi\right\rangle =\sum_{p,q=0}^{\infty}\sqrt{f_{p}\left(q\right)P_{q}}\left|p,p+q,q\right\rangle 
\end{equation}
and
\begin{equation}
\begin{split}\rho_{i} & =\left|\Psi\right\rangle \left\langle \Psi\right|\\
 & =\sum_{p_{1},q_{1}=0}^{\infty}\sum_{p_{2},q_{2}=0}^{\infty}\sqrt{f_{p_{1}}\left(q_{1}\right)P_{q_{1}}}\sqrt{f_{p_{2}}\left(q_{2}\right)P_{q_{2}}}\\
 &\times \left|p_{1},p_{1}+q_{1},q_{1}\right\rangle \left\langle p_{2},p_{2}+q_{2},q_{2}\right|
\end{split}
\end{equation}
first, the $\hat{M_{\mu}}\left(q\right)\left|\Psi\right\rangle $
is 

\begin{equation}
\begin{split}\hat{M_{\mu}}\left(q\right)\left|\Psi\right\rangle  & =\left[\sum_{n_{1}=q}^{\infty}\sqrt{C_{n_{1}}^{q}}\left(1-\mu\right)^{\frac{n_{1}-q}{2}}\mu^{\frac{q}{2}}\left|n_{1}-q\right\rangle _{q}\left\langle n_{1}\right|\right]\\
 & \times\left[\sum_{p_{1},q_{1}=0}^{\infty}\sqrt{f_{p_{1}}\left(q_{1}\right)P_{q_{1}}}\left|p_{1},p_{1}+q_{1},q_{1}\right\rangle \right]
\end{split}
\end{equation}
obviously $n_{1}=q_{1}$, so

\begin{equation}
\begin{split}\hat{M_{\mu}}\left(q\right)\left|\Psi\right\rangle  & =[\sum_{p_{1}=0}^{\infty}\sum_{q_{1}=q}^{\infty}\sqrt{C_{q_{1}}^{q}}\left(1-\mu\right)^{\frac{q_{1}-q}{2}}\mu^{\frac{q}{2}}\left|q_{1}-q\right\rangle _{q}\\
 & \times\sqrt{f_{p_{1}}\left(q_{1}\right)P_{q_{1}}}\left|p_{1},p_{1}+q_{1}\right\rangle ]
\end{split}
\end{equation}
second, the $\left\langle \Psi\right|\hat{M_{\mu}^{\dagger}}\left(q\right)$
is 

\begin{equation}
\begin{split}\left\langle \Psi\right|\hat{M_{\mu}^{\dagger}}\left(q\right) & =\left[\sum_{p_{2},q_{2}=0}^{\infty}\sqrt{f_{p_{2}}\left(q_{2}\right)P_{q_{2}}}\left\langle p_{2},p_{2}+q_{2},q_{2}\right|\right]\\
 & \times\left[\sum_{n_{2}=q}^{\infty}\sqrt{C_{n_{2}}^{q}}\left(1-\mu\right)^{\frac{n_{2}-q}{2}}\mu^{\frac{q}{2}}\left|n_{2}\right\rangle _{q}\left\langle n_{2}-q\right|\right]
\end{split}
\end{equation}
obviously $n_{2}=q_{2}$, so 

\begin{equation}
\begin{split}\left\langle \Psi\right|\hat{M_{\mu}^{\dagger}}\left(q\right) & =[\sum_{p_{2}=0}^{\infty}\sum_{q_{2}=q}^{\infty}\sqrt{f_{p_{2}}\left(q_{2}\right)P_{q_{2}}}\left\langle p_{2},p_{2}+q_{2}\right|\\
 & \times\sqrt{C_{q_{2}}^{q}}\left(1-\mu\right)^{\frac{q_{2}-q}{2}}\mu^{\frac{q}{2}}{}_{q}\left\langle q_{2}-q\right|]
\end{split}
\end{equation}
then, we have$\hat{M_{\mu}}\left(q\right)\rho\hat{M_{\mu}^{\dagger}}\left(q\right)$
\begin{equation}
\begin{split} & \hat{M_{\mu}}\left(q\right)\rho\hat{M_{\mu}^{\dagger}}\left(q\right)\\
= & \hat{M_{\mu}}\left(q\right)\left|\Psi\right\rangle \left\langle \Psi\right|\hat{M_{\mu}^{\dagger}}\left(q\right)\\
= & [\sum_{p_{1}=0}^{\infty}\sum_{q_{1}=q}^{\infty}\sqrt{C_{q_{1}}^{q}}\left(1-\mu\right){}^{\frac{q_{1}-q}{2}}\mu^{\frac{q}{2}}\left|q_{1}-q\right\rangle \\
 & \sqrt{f_{p_{1}}\left(q_{1}\right)P_{q_{1}}}\left|p_{1},p_{1}+q_{1}\right\rangle ]\\
\times & [\sum_{p_{2}=0}^{\infty}\sum_{q_{2}=q}^{\infty}\sqrt{f_{p_{2}}\left(q_{2}\right)P_{q_{2}}}\left\langle p_{2},p_{2}+q_{2}\right|\\
 & \sqrt{C_{q_{2}}^{q}}\left(1-\mu\right)^{\frac{q_{2}-q}{2}}\mu^{\frac{q}{2}}\left\langle q_{2}-q\right|]
\end{split}
\end{equation}
so we have 

\begin{equation}
Tr_{q}\left[\hat{M_{\mu}}\left(q\right)\rho\hat{M_{\mu}^{\dagger}}\left(q\right)\right]=\sum_{q_{3}=0}^{\infty}\left\langle q_{3}\right|\hat{M_{\mu}}\left(q\right)\left|\Psi\right\rangle \left\langle \Psi\right|\hat{M_{\mu}^{\dagger}}\left(q\right)\left|q_{3}\right\rangle 
\end{equation}
and we can find $q_{3}=q_{1}-q,q_{3}=q_{2}-q,q_{1}=q_{2}$ and let
$q_{1}=q_{2}=s$ finally, we have the conditional of state $p$ is 

\begin{equation}
\rho_{q}=\sum_{p_{1}=0}^{\infty}\sum_{,p_{2}=0}^{\infty}\sum_{s=q}^{\infty}\sqrt{f_{p_{1}}\left(s\right)f_{p_{2}}\left(s\right)}\frac{P_{s}C_{s}^{q}\left(1-\mu\right)^{s-q}\mu^{q}}{P_{\mu}\left(q\right)}\left|p_{1},p_{1}+s\right\rangle \left\langle p_{2},p_{2}+s\right|
\end{equation}
define 
\begin{equation}
\eta\left(s\right)=\frac{P_{s}C_{s}^{q}\left(1-\mu\right)^{s-q}\mu^{q}}{P_{\mu}\left(q\right)}=\frac{N_{m}^{s}}{\left(1+N_{m}\right)^{1+s}}\frac{\left(1+N_{m}\mu\right)^{1+q}}{N_{m}^{q}\mu^{q}}C_{s}^{q}\left(1-\mu\right)^{s-q}\mu^{q}
\end{equation}
, so we have 

\begin{equation}
\rho_{q}=\sum_{p_{1}=0}^{\infty}\sum_{,p_{2}=0}^{\infty}\sum_{s=q}^{\infty}\sqrt{f_{p_{1}}\left(s\right)f_{p_{2}}\left(s\right)}\eta\left(s\right)\left|p_{1},p_{1}+s\right\rangle \left\langle p_{2},p_{2}+s\right|
\end{equation}
also we can write $\rho_{q}$ in another way, because:

\begin{equation}
\left|\Psi_{s}\right\rangle =\sum_{p=0}^{\infty}\sqrt{f_{p}\left(s\right)}\left|p,p+s\right\rangle 
\end{equation}
then
\begin{equation}
\left|\Psi_{s}\right\rangle \left\langle \Psi_{s}\right|=\sum_{p_{1}=0}^{\infty}\sum_{,p_{2}=0}^{\infty}\sqrt{f_{p_{1}}\left(s\right)f_{p_{2}}\left(s\right)}\left|p_{1},p_{1}+s\right\rangle \left\langle p_{2},p_{2}+s\right|
\end{equation}
then 
\begin{equation}
\rho_{q}=\sum_{s=q}^{\infty}\eta\left(s\right)\left|\Psi_{s}\right\rangle \left\langle \Psi_{s}\right|
\end{equation}
its entanglement is $E_{N}^{\mu}\left(q\right)=\ln\left[1+2N\left(\rho_{q}^{\mathbf{T}}\right)\right]$
,where $N\left(\rho_{q}^{\mathbf{T}}\right)=\sum\left|\lambda_{i}\right|$,
and $\lambda_{i}$ are the negativity eigenvalues of $\rho_{q}^{\mathbf{T}}$.
$\rho_{q}^{\mathbf{T}}$ is the partial transpose of $\rho_{q}$ with
\begin{equation}
\rho_{q}^{\mathbf{T}}=\sum_{p_{1}=0}^{\infty}\sum_{,p_{2}=0}^{\infty}\sum_{s=q}^{\infty}\sqrt{f_{p_{1}}\left(s\right)f_{p_{2}}\left(s\right)}\eta\left(s\right)\left|p_{1},p_{2}+s\right\rangle \left\langle p_{2},p_{1}+s\right|
\end{equation}

\subsection*{Supplement[D]: Diagonalize the density matrix}

the partial transpose of the density matrix is
\begin{equation}
\rho_{q}^{\mathbf{T}}=\sum_{p_{1}=0}^{\infty}\sum_{,p_{2}=0}^{\infty}\sum_{s=q}^{\infty}\sqrt{f_{p_{1}}\left(s\right)f_{p_{2}}\left(s\right)}\eta\left(s\right)\left|p_{1},p_{2}+s\right\rangle \left\langle p_{2},p_{1}+s\right|
\end{equation}
We let $Q=p_{1}+p_{2}+s$$\left(Q\geq q\right)$, and define 
\begin{equation}
F\left[p_{1},p_{2},s\right]=\sqrt{f_{p_{1}}\left(s\right)f_{p_{2}}\left(s\right)}\eta\left(s\right)
\end{equation}
Then we get 

when$Q=q$

\begin{tabular}{|c|c|}
\hline 
$\left|p_{1},p_{2}+s\right\rangle \left\langle p_{2},p_{1}+s\right|$ & $\left\langle 0,q\right|$\tabularnewline
\hline 
\hline 
$\left|0,q\right\rangle $ & $\begin{array}{c}
p_{1}=0\\
p_{2}=0\\
s=q
\end{array}$\tabularnewline
\hline 
\end{tabular}

\[
Q\left[q\right]=F\left[0,0,q\right]
\]

when $Q=q+1$

\begin{tabular}{|c|c|c|}
\hline 
$\left|p_{1},p_{2}+s\right\rangle \left\langle p_{2},p_{1}+s\right|$ & $\left\langle 0,q+1\right|$ & $\left\langle 1,q\right|$\tabularnewline
\hline 
\hline 
$\left|1,q\right\rangle $ & $\begin{array}{c}
p_{1}=1\\
p_{2}=0\\
s=q
\end{array}$ & 0\tabularnewline
\hline 
$\left|0,q+1\right\rangle $ & $\begin{array}{c}
p_{1}=0\\
p_{2}=0\\
s=q+1
\end{array}$ & $\begin{array}{c}
p_{1}=0\\
p_{2}=1\\
s=q
\end{array}$\tabularnewline
\hline 
\end{tabular}

\[
Q\left[q+1\right]=\left[\begin{array}{cc}
F\left[1,0,q\right] & 0\\
F\left[0,0,q+1\right] & F\left[0,1,q\right]
\end{array}\right]
\]

when $Q=q+2$

\begin{tabular}{|c|c|c|c|}
\hline 
$\left|p_{1},p_{2}+s\right\rangle \left\langle p_{2},p_{1}+s\right|$ & $\left\langle 0,q+2\right|$ & $\left\langle 1,q+1\right|$ & $\left\langle 2,q\right|$\tabularnewline
\hline 
\hline 
$\left|2,q\right\rangle $ & $\begin{array}{c}
p_{1}=2\\
p_{2}=0\\
s=q
\end{array}$ & $0$ & 0\tabularnewline
\hline 
$\left|1,q+1\right\rangle $ & $\begin{array}{c}
p_{1}=1\\
p_{2}=0\\
s=q+1
\end{array}$ & $\begin{array}{c}
p_{1}=1\\
p_{2}=1\\
s=q
\end{array}$ & 0\tabularnewline
\hline 
$\left|0,q+2\right\rangle $ & $\begin{array}{c}
p_{1}=0\\
p_{2}=0\\
s=q+2
\end{array}$ & $\begin{array}{c}
p_{1}=0\\
p_{2}=1\\
s=q+1
\end{array}$ & $\begin{array}{c}
p_{1}=0\\
p_{2}=2\\
s=q
\end{array}$\tabularnewline
\hline 
\end{tabular}

\[
Q\left[q+2\right]=\left[\begin{array}{ccc}
F\left[2,0,q\right] & 0 & 0\\
F\left[1,0,q+1\right] & F\left[1,1,q\right] & 0\\
F\left[0,0,q+2\right] & F\left[0,1,q+1\right] & F\left[0,2,q\right]
\end{array}\right]
\]

so when $Q=q+n$

\begin{tabular}{|c|c|c|c|c|}
\hline 
$\left|p_{1},p_{2}+s\right\rangle \left\langle p_{2},p_{1}+s\right|$ & $\left\langle 0,q+n\right|$ & $...$ & $\left\langle n-1,q+1\right|$ & $\left\langle n,q\right|$\tabularnewline
\hline 
\hline 
$\left|n,q\right\rangle $ & $\begin{array}{c}
p_{1}=n\\
p_{2}=0\\
s=q
\end{array}$ & $...$ & 0 & 0\tabularnewline
\hline 
$\left|n-1,q+1\right\rangle $ & $\begin{array}{c}
p_{1}=n-1\\
p_{2}=0\\
s=q+1
\end{array}$ & $...$ & $0$ & 0\tabularnewline
\hline 
$...$ & $...$ & $...$ & $...$ & $...$\tabularnewline
\hline 
$\left|0,q+n\right\rangle $ & $\begin{array}{c}
p_{1}=0\\
p_{2}=0\\
s=q+n
\end{array}$ & $...$ & $\begin{array}{c}
p_{1}=0\\
p_{2}=n-1\\
s=q+1
\end{array}$ & $\begin{array}{c}
p_{1}=0\\
p_{2}=n\\
s=q
\end{array}$\tabularnewline
\hline 
\end{tabular}

\[
Q\left[q+n\right]=\left[\begin{array}{cccc}
F\left[n,0,q\right] & ... & 0 & 0\\
F\left[n-1,0,q+1\right] & ... & 0 & 0\\
... & ... & ... & ...\\
F\left[0,0,q+n\right] & ... & F\left[0,n-1,q+1\right] & F\left[0,n,q\right]
\end{array}\right]
\]
In this way we can diagonalize the density matrix with

\begin{equation}
\hat{\rho}_{q}^{\mathbf{PT}}=\left[\begin{array}{ccccc}
Q\left[q\right]\\
 & Q\left[q+1\right]\\
 &  & ...\\
 &  &  & Q\left[q+n\right]\\
 &  &  &  & ...
\end{array}\right]
\end{equation}

\subsection*{Supplement[E]: Limit Case of $\mu$ }

If $\mu=1$ only $s=q$ is valid, so $\eta\left(q\right)=1$ ,

\begin{equation}
\rho_{q}=\sum_{p_{1}=0}^{\infty}\sum_{,p_{2}=0}^{\infty}\sqrt{f_{p_{1}}\left(q\right)f_{p_{2}}\left(q\right)}\left|p_{1},p_{1}+q\right\rangle \left\langle p_{2},p_{2}+q\right|
\end{equation}
\begin{equation}
\left|\Psi_{q}\right\rangle \left\langle \Psi_{q}\right|=\sum_{p_{1}p_{2}=0}^{\infty}\sqrt{f_{p_{1}}\left(q\right)f_{p_{2}}\left(q\right)}\left|p_{1},p_{1}+q\right\rangle \left\langle p_{2},p_{2}+q\right|
\end{equation}
So

\begin{equation}
\rho_{q}=\left|\Psi_{q}\right\rangle \left\langle \Psi_{q}\right|
\end{equation}
When $\mu\longrightarrow1$, imperfect measurement will turn to be
perfect measurement.

If $\mu\longrightarrow0$ , $\eta\left(s\right)=\frac{N_{m}^{s}C_{s}^{q}}{N_{m}^{q}\left(1+N_{m}\right)^{1+s}}$,
\begin{equation}
\rho_{q}=\sum_{p_{1}=0}^{\infty}\sum_{,p_{2}=0}^{\infty}\sum_{s=q}^{\infty}\sqrt{f_{p_{1}}\left(s\right)f_{p_{2}}\left(s\right)}\frac{N_{m}^{s}C_{s}^{q}}{N_{m}^{q}\left(1+N_{m}\right)^{1+s}}\left|p_{1},p_{1}+s\right\rangle \left\langle p_{2},p_{2}+s\right|
\end{equation}

\begin{equation}
\rho_{q}=\sum_{s=q}^{\infty}\frac{N_{m}^{s}C_{s}^{q}}{N_{m}^{q}\left(1+N_{m}\right)^{1+s}}\left|\Psi_{s}\right\rangle \left\langle \Psi_{s}\right|
\end{equation}
from a physical standpoint it is necessary to make $q=0$.

\begin{equation}
\rho_{0}=\sum_{s=0}^{\infty}\frac{N_{m}^{s}}{\left(1+N_{m}\right)^{1+s}}\left|\Psi_{s}\right\rangle \left\langle \Psi_{s}\right|
\end{equation}
to make $\rho_{0}$ more clear, we may calculate this $Tr_{q}\left[\left|\Psi\right\rangle \left\langle \Psi\right|\right]$
\begin{equation}
Tr_{q}\left[\left|\Psi\right\rangle \left\langle \Psi\right|\right]=\sum_{s=0}^{\infty}\left\langle s\left|\Psi\right\rangle \left\langle \Psi\right|s\right\rangle 
\end{equation}
\begin{equation}
\left\langle s\mid\Psi\right\rangle =\sum_{p_{1},q_{1}=0}^{\infty}\sqrt{f_{p_{1}}\left(q_{1}\right)P_{q_{1}}}\left\langle s\mid p_{1},p_{1}+q_{1},q_{1}\right\rangle 
\end{equation}
\begin{equation}
\left\langle \Psi\mid s\right\rangle =\sum_{p_{2},q_{2}=0}^{\infty}\sqrt{f_{p_{2}}\left(q_{2}\right)P_{q_{2}}}\left\langle p_{2},p_{2}+q_{2},q_{2}\mid s\right\rangle 
\end{equation}
the condition is $s=q_{1}=q_{2}$
\begin{equation}
Tr_{q}\left[\left|\Psi\right\rangle \left\langle \Psi\right|\right]=\sum_{s=0}^{\infty}\sum_{p_{1},p_{2}=0}^{\infty}\sqrt{f_{p_{1}}\left(s\right)P_{s}}\sqrt{f_{p_{2}}\left(s\right)P_{s}}\left|p_{1},p_{1}+s\right\rangle \left\langle p_{2},p_{2}+s\right|
\end{equation}
Simplify it, we can get 
\begin{equation}
Tr_{q}\left[\left|\Psi\right\rangle \left\langle \Psi\right|\right]=\sum_{s=0}^{\infty}\frac{N_{m}^{s}}{\left(1+N_{m}\right)^{1+s}}\left|\Psi_{s}\right\rangle \left\langle \Psi_{s}\right|
\end{equation}
The  $Tr_{q}\left[\left|\Psi\right\rangle \left\langle \Psi\right|\right]$
and $\rho_{0}$ are the same
\begin{equation}
\rho_{0}=Tr_{q}\left[\left|\Psi\right\rangle \left\langle \Psi\right|\right].
\end{equation}

\subsection*{Supplement[F]: Entanglement Calculation after Imperfect Measuement}

The state after measurement is
\begin{equation}
\rho_{q}=\sum_{p_{1}=0}^{\infty}\sum_{,p_{2}=0}^{\infty}\sum_{s=q}^{\infty}\sqrt{f_{p_{1}}\left(s\right)f_{p_{2}}\left(s\right)}\eta\left(s\right)\left|p_{1},p_{1}+s\right\rangle \left\langle p_{2},p_{2}+s\right|
\end{equation}
with $\eta\left(s\right)=\frac{N_{m}^{s}}{\left(1+N_{m}\right)^{1+s}}\frac{\left(1+N_{m}\mu\right)^{1+q}}{N_{m}^{q}\mu^{q}}C_{s}^{q}\left(1-\mu\right)^{s-q}\mu^{q}$.
The partial transpose is 
\begin{equation}
\rho_{q}^{\mathbf{T}}=\sum_{p_{1}=0}^{\infty}\sum_{,p_{2}=0}^{\infty}\sum_{s=q}^{\infty}\sqrt{f_{p_{1}}\left(s\right)f_{p_{2}}\left(s\right)}\eta\left(s\right)\left|p_{1},p_{2}+s\right\rangle \left\langle p_{2},p_{1}+s\right|
\end{equation}
The sum of $s$ is from $q$ to $+\infty$ , it is difficult to calculate
the negativity eigenvalues of $\rho_{q}^{\mathbf{T}}$ directly. Here
we use perturbation theory, and we just consider two term $s=q$ ,
$s=q+1$ . So we rewrite it

\begin{equation}
\begin{split}\rho_{q}^{\mathbf{T}} & =\sum_{p_{1}=0}^{\infty}\sum_{,p_{2}=0}^{\infty}\sqrt{f_{p_{1}}\left(q\right)f_{p_{2}}\left(q\right)}\eta\left(q\right)\left|p_{1},p_{2}+q\right\rangle \left\langle p_{2},p_{1}+q\right|\\
 & +\sum_{i=0}^{\infty}\sum_{,j=0}^{\infty}\sqrt{f_{i}\left(q+1\right)f_{j}\left(q+1\right)}\eta\left(q+1\right)\left|i,j+q+1\right\rangle \left\langle j,i+q+1\right|
\end{split}
\end{equation}
and define
\begin{equation}
H_{0}=\sum_{p_{1}=0}^{\infty}\sum_{,p_{2}=0}^{\infty}\sqrt{f_{p_{1}}\left(q\right)f_{p_{2}}\left(q\right)}\eta\left(q\right)\left|p_{1},p_{2}+q\right\rangle \left\langle p_{2},p_{1}+q\right|
\end{equation}
\begin{equation}
H_{1}=\sum_{i=0}^{\infty}\sum_{,j=0}^{\infty}\sqrt{f_{i}\left(q+1\right)f_{j}\left(q+1\right)}\eta\left(q+1\right)\left|i,j+q+1\right\rangle \left\langle j,i+q+1\right|
\end{equation}
so
\begin{equation}
\rho_{q}^{\mathbf{T}}=H_{0}+H_{1}
\end{equation}
for $H_{0}$ , we have the eigenvalues and eigenvectors
\begin{equation}
\begin{array}{ccc}
condition & eigenvectors & eigenvalues\\
p_{1}=p_{2}=p & \left|\alpha\right\rangle =\left|p,p+q\right\rangle  & f_{p}\left(q\right)\eta\left(q\right)\\
p_{1}<p_{2} & \left|\beta_{+}\right\rangle =\frac{\left|p_{1},p_{2}+q\right\rangle +\left|p_{2},p_{1}+q\right\rangle }{\sqrt{2}} & +\sqrt{f_{p_{1}}\left(q\right)f_{p_{2}}\left(q\right)}\eta\left(q\right)\\
p_{1}<p_{2} & \left|\beta_{-}\right\rangle =\frac{\left|p_{1},p_{2}+q\right\rangle -\left|p_{2},p_{1}+q\right\rangle }{\sqrt{2}} & -\sqrt{f_{p_{1}}\left(q\right)f_{p_{2}}\left(q\right)}\eta\left(q\right)
\end{array}
\end{equation}
and we can calculate entanglement in this way

\begin{equation}
E_{N}^{\mu}\left(q\right)=\ln\left(1+2\sum_{\beta_{-}}\left|E_{\beta_{-}}\right|\right)
\end{equation}
with
\begin{equation}
\begin{split}E_{\beta_{-}}= & E_{\beta_{-}}^{\left(0\right)}+E_{\beta_{-}}^{\left(1\right)}+E_{\beta_{-}}^{\left(2\right)}\\
E_{\beta_{-}}^{\left(0\right)}= & -\sqrt{f_{p_{1}}\left(q\right)f_{p_{2}}\left(q\right)}\eta\left(q\right)\\
E_{\beta_{-}}^{\left(1\right)}= & \left\langle \beta_{-}\left|H_{1}\right|\beta_{-}\right\rangle \\
E_{\beta_{-}}^{\left(2\right)}= & \sum_{n\neq\beta_{-}}\frac{\left(\left\langle \beta_{-}\left|H_{1}\right|n\right\rangle \right)^{2}}{E_{\beta_{-}}^{\left(0\right)}-E_{n}^{0}}
\end{split}
\end{equation}
and in this article, because $E_{\beta_{-}}^{\left(0\right)}<0$ ,
$\left|E_{\beta_{-}}^{\left(0\right)}\right|\gg\left|E_{\beta_{-}}^{\left(1\right)}\right|$,
$\left|E_{\beta_{-}}^{\left(0\right)}\right|\gg\left|E_{\beta_{-}}^{\left(2\right)}\right|$,
so we have 
\begin{equation}
\left|E_{\beta_{-}}\right|=\left|E_{\beta_{-}}^{\left(0\right)}+E_{\beta_{-}}^{\left(1\right)}+E_{\beta_{-}}^{\left(2\right)}\right|=-E_{\beta_{-}}^{\left(0\right)}-E_{\beta_{-}}^{\left(1\right)}-E_{\beta_{-}}^{\left(2\right)}
\end{equation}
\begin{equation}
\sum_{\beta_{-}}\left|E_{\beta_{-}}\right|=\sum_{\beta_{-}}\left(-E_{\beta_{-}}^{\left(0\right)}-E_{\beta_{-}}^{\left(1\right)}-E_{\beta_{-}}^{\left(2\right)}\right)
\end{equation}
\begin{equation}
E_{N}^{\mu}\left(q\right)=\ln\left[1+2\sum_{\beta_{-}}\left(-E_{\beta_{-}}^{\left(0\right)}-E_{\beta_{-}}^{\left(1\right)}-E_{\beta_{-}}^{\left(2\right)}\right)\right]
\end{equation}
So the next mission is to calculate $\sum_{\beta_{-}}E_{\beta_{-}}^{\left(0\right)},\sum_{\beta_{-}}E_{\beta_{-}}^{\left(1\right)},\sum_{\beta_{-}}E_{\beta_{-}}^{\left(2\right)}$.

\subsection*{Supplement[G]: Zero Order and First Order Calculation }

For the zero order calculation , it is easy.

\begin{equation}
\begin{split}\sum_{\beta_{-}}-E_{\beta_{-}}^{\left(0\right)} & =\sum_{p_{1}<p_{2}}^{\infty}\left(-E_{\beta_{-}}^{\left(0\right)}\right)\\
 & =\sum_{p_{1}<p_{2}}^{\infty}\sqrt{f_{p_{1}}\left(q\right)f_{p_{2}}\left(q\right)}\eta\left(q\right)\\
 & =\frac{1}{2}\sum_{p_{1}\neq p_{2}}^{\infty}\sqrt{f_{p_{1}}\left(q\right)f_{p_{2}}\left(q\right)}\eta\left(q\right)\\
 & =\frac{1}{2}\sum_{p_{1},p_{2}}^{\infty}\sqrt{f_{p_{1}}\left(q\right)f_{p_{2}}\left(q\right)}\eta\left(q\right)-\frac{1}{2}\sum_{p_{1}=p_{2}}^{\infty}\sqrt{f_{p_{1}}\left(q\right)f_{p_{2}}\left(q\right)}\eta\left(q\right)\\
 & =\frac{1}{2}\left(\sum_{p=0}^{\infty}\sqrt{f_{p}\left(q\right)}\right)^{2}\eta\left(q\right)-\frac{1}{2}\eta\left(q\right)
\end{split}
\end{equation}
For the first order calculation, the following formulas may be useful

\begin{equation}
\begin{split} & \langle\beta_{-}\left|i,j+q+1\right\rangle \left\langle j,i+q+1\right|\\
= & \frac{\left\langle p_{1},p_{2}+q\right|-\left\langle p_{2},p_{1}+q\right|}{\sqrt{2}}\left|i,j+q+1\right\rangle \left\langle j,i+q+1\right|\\
= & \frac{\left\langle p_{2}-1,p_{1}+q+1\right|}{\sqrt{2}}-\frac{\left\langle p_{1}-1,p_{2}+q+1\right|}{\sqrt{2}}
\end{split}
\end{equation}
 
\begin{equation}
\begin{split} & \left\langle \beta_{-}\right|H_{1}\\
= & \frac{\left\langle p_{1},p_{2}+q\right|-\left\langle p_{2},p_{1}+q\right|}{\sqrt{2}}H_{1}\\
= & \eta\left(q+1\right)\sqrt{f_{p_{2}-1}\left(q+1\right)f_{p_{1}}\left(q+1\right)}\frac{\left\langle p_{2}-1,p_{1}+q+1\right|}{\sqrt{2}}\\
- & \eta\left(q+1\right)\sqrt{f_{p_{1}-1}\left(q+1\right)f_{p_{2}}\left(q+1\right)}\frac{\left\langle p_{1}-1,p_{2}+q+1\right|}{\sqrt{2}}
\end{split}
\end{equation}
we calculate$E_{\beta_{-}}^{\left(1\right)}=\left\langle \beta_{-}\left|H_{1}\right|\beta_{-}\right\rangle $:

\begin{equation}
\begin{split} & \left\langle \beta_{-}\left|H_{1}\right|\beta_{-}\right\rangle \\
= & \eta\left(q+1\right)\sqrt{f_{p_{2}-1}\left(q+1\right)f_{p_{1}}\left(q+1\right)}\frac{\left\langle p_{2}-1,p_{1}+q+1\right|}{\sqrt{2}}\frac{\left|p_{1},p_{2}+q\right\rangle -\left|p_{2},p_{1}+q\right\rangle }{\sqrt{2}}\\
- & \eta\left(q+1\right)\sqrt{f_{p_{1}-1}\left(q+1\right)f_{p_{2}}\left(q+1\right)}\frac{\left\langle p_{1}-1,p_{2}+q+1\right|}{\sqrt{2}}\frac{\left|p_{1},p_{2}+q\right\rangle -\left|p_{2},p_{1}+q\right\rangle }{\sqrt{2}}\\
= & \frac{\eta\left(q+1\right)}{2}\left(\sqrt{f_{p_{2}-1}\left(q+1\right)f_{p_{1}}\left(q+1\right)}\delta_{p_{2},p_{1}+1}+\sqrt{f_{p_{1}-1}\left(q+1\right)f_{p_{2}}\left(q+1\right)}\delta_{p_{1},p_{2}+1}\right)\\
= & \frac{\eta\left(q+1\right)}{2}\left(f_{p_{1}}\left(q+1\right)\delta_{p_{2},p_{1}+1}+f_{p_{2}}\left(q+1\right)\delta_{p_{1},p_{2}+1}\right)
\end{split}
\end{equation}
\begin{equation}
E_{\beta_{-}}^{\left(1\right)}=\langle\left\langle \beta_{-}\left|H_{1}\right|\beta_{-}\right\rangle =\frac{\eta\left(q+1\right)}{2}\left(f_{p_{1}}\left(q+1\right)\delta_{p_{2},p_{1}+1}+f_{p_{2}}\left(q+1\right)\delta_{p_{1},p_{2}+1}\right)
\end{equation}
so 
\begin{equation}
\begin{split}\sum_{\beta_{-}}-E_{\beta_{-}}^{\left(1\right)} & =\sum_{p_{1}<p_{2}}^{\infty}\left(-E_{\beta_{-}}^{\left(1\right)}\right)\\
 & =\frac{1}{2}\sum_{p_{1}\neq p_{2}}^{\infty}\left(-E_{\beta_{-}}^{\left(1\right)}\right)\\
 & =-\frac{1}{2}\sum_{p_{1}\neq p_{2}}^{\infty}\frac{\eta\left(q+1\right)}{2}\left(f_{p_{1}}\left(q+1\right)\delta_{p_{2},p_{1}+1}+f_{p_{2}}\left(q+1\right)\delta_{p_{1},p_{2}+1}\right)\\
 & =-\frac{\eta\left(q+1\right)}{4}\left(\sum_{p_{1}\neq p_{2}}^{\infty}f_{p_{1}}\left(q+1\right)\delta_{p_{2},p_{1}+1}+\sum_{p_{1}\neq p_{2}}^{\infty}f_{p_{2}}\left(q+1\right)\delta_{p_{1},p_{2}+1}\right)\\
 & =-\frac{\eta\left(q+1\right)}{2}
\end{split}
\end{equation}
If we just consider zero order and first order ,
\begin{equation}
E_{\beta_{-}}=E_{\beta_{-}}^{\left(0\right)}+E_{\beta_{-}}^{\left(1\right)}
\end{equation}
\begin{equation}
\begin{split}\sum_{\beta_{-}}\left|E_{\beta_{-}}\right| & =\sum_{\beta_{-}}\left(-E_{\beta_{-}}^{\left(0\right)}\right)+\sum_{\beta_{-}}\left(-E_{\beta_{-}}^{\left(1\right)}\right)\\
 & =\frac{1}{2}\left(\sum_{p=0}^{\infty}\sqrt{f_{p}\left(q\right)}\right)^{2}\eta\left(q\right)-\frac{1}{2}\eta\left(q\right)-\frac{\eta\left(q+1\right)}{2}
\end{split}
\end{equation}
So we get the expression of $E_{N}^{\mu}\left(q\right)$
\begin{equation}
\begin{split}E_{N}^{\mu}\left(q\right) & =\ln\left[1+2\sum_{\beta_{-}}\left(-E_{\beta_{-}}^{\left(0\right)}-E_{\beta_{-}}^{\left(1\right)}\right)\right]\\
 & =\ln\left[1+\left(\sum_{p=0}^{\infty}\sqrt{f_{p}\left(q\right)}\right)^{2}\eta\left(q\right)-\eta\left(q\right)-\eta\left(q+1\right)\right]
\end{split}
\end{equation}
Use the accurate expression of $\eta\left(s\right)$ ,we get :
\begin{equation}
\eta\left(q\right)=\left(\frac{1+N_{m}\mu}{1+N_{m}}\right)^{1+q}=\left(1+\frac{N_{m}\left(\mu-1\right)}{1+N_{m}}\right)^{1+q}=\left(1-\varepsilon\right)^{1+q}
\end{equation}
\begin{equation}
\eta\left(q+1\right)=\frac{\left(q+1\right)\left(1-\mu\right)N_{m}}{\left(1+N_{m}\right)}\left(\frac{1+N_{m}\mu}{1+N_{m}}\right)^{1+q}=\left(q+1\right)\varepsilon\left(1-\varepsilon\right)^{1+q}
\end{equation}
here we define:$\varepsilon=\frac{\left(1-\mu\right)N_{m}}{\left(1+N_{m}\right)}$
, and use the approximation formula $\left(1+x\right)^{n}\approx1+nx$.
\begin{equation}
\left(1-\varepsilon\right)^{1+q}\approx1-\left(q+1\right)\varepsilon
\end{equation}
We throw away some things such as $O\left[\left(1-\mu\right)^{2}\right]$,
we can obtain 
\begin{equation}
\begin{array}{ccc}
\eta\left(q\right)\approx1-\left(q+1\right)\varepsilon & , & \eta\left(q+1\right)\approx\left(q+1\right)\varepsilon\end{array}
\end{equation}
We can get a simple expression of $E_{N}^{\mu}\left(q\right)$
\begin{equation}
E_{N}^{\mu}\left(q\right)=\ln\left[\left(\sum_{p}^{\infty}\sqrt{f_{p}\left(q\right)}\right)^{2}\left(1-\left(q+1\right)\varepsilon\right)\right]
\end{equation}
\begin{equation}
E_{N}^{\mu}\left(q\right)=2\ln\sum_{p}^{\infty}\sqrt{f_{p}\left(q\right)}+ln\left(1-\left(q+1\right)\varepsilon\right)
\end{equation}
make a further approximation
\begin{equation}
E_{N}^{\mu}\left(q\right)=2\ln\sum_{p}^{\infty}\sqrt{f_{p}\left(q\right)}-\left(q+1\right)\varepsilon
\end{equation}
with $\varepsilon=\frac{\left(1-\mu\right)N_{m}}{\left(1+N_{m}\right)}$
. In the perfect measurement, we have 
\begin{equation}
E_{N}\left(q\right)=2\ln\sum_{p}^{\infty}\sqrt{f_{p}\left(q\right)}
\end{equation}
So we can get the entanglement of imperfcet measurement after first
order perturbation approximation 
\begin{equation}
E_{N}^{\mu}\left(q\right)=E_{N}\left(q\right)-\left(q+1\right)\varepsilon
\end{equation}

\subsection*{Supplement[H]: Second Order Calculation }

The second order is$E_{\beta_{-}}^{\left(2\right)}=\sum_{n\neq\beta_{-}}\frac{\left(\left\langle \beta_{-}\left|H_{1}\right|n\right\rangle \right)^{2}}{E_{\beta_{-}}^{\left(0\right)}-E_{n}^{0}}$.
$n$ is the eigenvectors of $H_{0}$ with the restricted condition
$n\neq\beta_{-}$ . We just care $\sum_{\beta_{-}}E_{\beta_{-}}^{\left(2\right)}$
. First,
\begin{equation}
\begin{split}E_{\beta_{-}}^{\left(2\right)} & =\sum_{n\neq\beta_{-}}\frac{\left(\left\langle \beta_{-}\left|H_{1}\right|n\right\rangle \right)^{2}}{E_{\beta_{-}}^{\left(0\right)}-E_{n}^{0}}\\
 & =\sum_{\alpha_{-}\neq\beta_{-}}\frac{\left(\left\langle \beta_{-}\left|H_{1}\right|\alpha_{-}\right\rangle \right)^{2}}{E_{\beta_{-}}^{\left(0\right)}-E_{\alpha_{-}}^{0}}+\sum_{\alpha_{+}}\frac{\left(\left\langle \beta_{-}\left|H_{1}\right|\alpha_{+}\right\rangle \right)^{2}}{E_{\beta_{-}}^{\left(0\right)}-E_{\alpha_{+}}^{0}}+\sum_{\alpha}\frac{\left(\left\langle \beta_{-}\left|H_{1}\right|\alpha\right\rangle \right)^{2}}{E_{\beta_{-}}^{\left(0\right)}-E_{\alpha}^{0}}
\end{split}
\end{equation}
and
\begin{equation}
\begin{split}\sum_{\beta_{-}}E_{\beta_{-}}^{\left(2\right)}= & \sum_{\beta_{-}}\sum_{\alpha_{-}\neq\beta_{-}}\frac{\left(\left\langle \beta_{-}\left|H_{1}\right|\alpha_{-}\right\rangle \right)^{2}}{E_{\beta_{-}}^{\left(0\right)}-E_{\alpha_{-}}^{0}}+\sum_{\beta_{-}}\sum_{\alpha_{+}}\frac{\left(\left\langle \beta_{-}\left|H_{1}\right|\alpha_{+}\right\rangle \right)^{2}}{E_{\beta_{-}}^{\left(0\right)}-E_{\alpha_{+}}^{0}}\\
 & +\sum_{\beta_{-}}\sum_{\alpha}\frac{\left(\left\langle \beta_{-}\left|H_{1}\right|\alpha\right\rangle \right)^{2}}{E_{\beta_{-}}^{\left(0\right)}-E_{\alpha}^{0}}
\end{split}
\end{equation}
obvious
\begin{equation}
\sum_{\beta_{-}}\sum_{\alpha_{-}\neq\beta_{-}}\frac{\left(\left\langle \beta_{-}\left|H_{1}\right|\alpha_{-}\right\rangle \right)^{2}}{E_{\beta_{-}}^{\left(0\right)}-E_{\alpha_{-}}^{0}}=0
\end{equation}
so we only care
\begin{equation}
\sum_{k_{-}}E_{k_{-}}^{\left(2\right)}=\sum_{\beta_{-}}\sum_{\alpha_{+}}\frac{\left(\left\langle \beta_{-}\left|H_{1}\right|\alpha_{+}\right\rangle \right)^{2}}{E_{\beta_{-}}^{\left(0\right)}-E_{\alpha_{+}}^{0}}+\sum_{\beta_{-}}\sum_{\alpha}\frac{\left(\left\langle \beta_{-}\left|H_{1}\right|\alpha\right\rangle \right)^{2}}{E_{\beta_{-}}^{\left(0\right)}-E_{\alpha}^{0}}
\end{equation}
also, define a expression to simplify calculation
\begin{equation}
g\left(p_{1},p_{2}\right)=\frac{f_{p_{2}}\left(q+1\right)f_{p_{1}}\left(q+1\right)}{\sqrt{f_{p_{1}}\left(q\right)f_{p_{2}+1}\left(q\right)}+\sqrt{f_{p_{1}+1}\left(q\right)f_{p_{2}}\left(q\right)}}
\end{equation}
also because $f_{p}\left(q\right)=C_{p+q}^{p}\zeta^{p}\left(1-\zeta\right)^{1+q}$
we have a lot of different expressions of $g\left(p_{1},p_{2}\right)$
and we just write them for further use.
\begin{equation}
g\left(p_{1},p_{2}\right)=\frac{C_{p_{2}+q+1}^{p_{2}}C_{p_{1}+q+1}^{p_{1}}}{\sqrt{C_{p_{1}+q}^{p_{1}}C_{p_{2}+1+q}^{p_{2}+1}}+\sqrt{C_{p_{1}+1+q}^{p_{1}+1}C_{p_{2}+q}^{p_{2}}}}\frac{\zeta^{p_{1}+p_{2}}\left(1-\zeta\right)^{4+2q}}{\sqrt{\zeta^{p_{1}+p_{2}+1}\left(1-\zeta\right)^{2+2q}}}
\end{equation}
\begin{equation}
g\left(p_{1},p_{2}\right)=\frac{C_{p_{2}+q}^{p_{2}}C_{p_{1}+q}^{p_{1}}\frac{p_{2}+q+1}{q+1}\frac{p_{1}+q+1}{q+1}}{\sqrt{C_{p_{1}+q}^{p_{1}}C_{p_{2}+q}^{p_{2}}\frac{p_{2}+1+q}{p_{2}+1}}+\sqrt{C_{p_{1}+q}^{p_{1}}C_{Pp_{2}+q}^{p_{2}}\frac{p_{1}+1+q}{p_{1}+1}}}\frac{\sqrt{\zeta^{p_{1}+p_{2}}}\left(1-\zeta\right)^{3+q}}{\sqrt{\zeta}}
\end{equation}
\begin{equation}
g\left(p_{1},p_{2}\right)=\sqrt{C_{p_{2}+q}^{p_{2}}C_{p_{1}+q}^{p_{1}}}\frac{\sqrt{\zeta^{p_{1}+p_{2}}\left(1-\zeta\right)^{2+2q}}\left(1-\zeta\right)^{2}}{\sqrt{\zeta}}\frac{\frac{p_{2}+q+1}{q+1}\frac{p_{1}+q+1}{q+1}}{\sqrt{\frac{p_{2}+1+q}{p_{2}+1}}+\sqrt{\frac{p_{1}+1+q}{p_{1}+1}}}
\end{equation}
\begin{equation}
g\left(p_{1},p_{2}\right)=\sqrt{f_{p_{1}}\left(q\right)f_{p_{2}}\left(q\right)}\frac{\left(1-\zeta\right)^{2}}{\sqrt{\zeta}\left(q+1\right)^{2}}\frac{\left(p_{1}+q+1\right)\left(p_{2}+q+1\right)}{\sqrt{1+\frac{q}{p_{1}+1}}+\sqrt{1+\frac{q}{p_{2}+1}}}
\end{equation}
and we first care $\sum_{\beta_{-}}\sum_{\alpha}\frac{\left(\left\langle \beta_{-}\left|H_{1}\right|\alpha\right\rangle \right)^{2}}{E_{\beta_{-}}^{\left(0\right)}-E_{\alpha}^{0}}$
, and the following formulas may be useful 
\begin{equation}
\begin{split} & \left\langle \beta_{-}\right|H_{1}\left|p,p+q\right\rangle \\
 & \frac{\left\langle p_{1},p_{2}+q\right|-\left\langle p_{2},p_{1}+q\right|}{\sqrt{2}}H_{1}\left|p,p+q\right\rangle \\
= & \frac{\eta\left(q+1\right)}{\sqrt{2}}\sqrt{f_{p_{2}-1}\left(q+1\right)f_{p_{1}}\left(q+1\right)}\delta_{p,p_{1}+1}\delta_{p_{2},p_{1}+2}\\
- & \frac{\eta\left(q+1\right)}{\sqrt{2}}\sqrt{f_{p_{1}-1}\left(q+1\right)f_{p_{2}}\left(q+1\right)}\delta_{p,p_{1}-1}\delta_{p_{2},p_{1}-2}
\end{split}
\end{equation}
because $p_{1}<p_{2}$ 
\begin{equation}
\left\langle \beta_{-}\right|H_{1}\left|p,p+q\right\rangle =\frac{\eta\left(q+1\right)}{\sqrt{2}}\sqrt{f_{p_{2}-1}\left(q+1\right)f_{p_{1}}\left(q+1\right)}\delta_{p,p_{1}+1}\delta_{p_{2},p_{1}+2}
\end{equation}
\begin{equation}
\left\langle \beta_{-}\right|H_{1}\left|p,p+q\right\rangle ^{2}=\frac{\eta^{2}\left(q+1\right)}{2}f_{p_{2}-1}\left(q+1\right)f_{p_{1}}\left(q+1\right)\delta_{p,p_{1}+1}\delta_{p_{2},p_{1}+2}
\end{equation}
so
\begin{equation}
\begin{split}M_{1} & =\frac{\frac{\eta^{2}\left(q+1\right)}{2}f_{p_{2}-1}\left(q+1\right)f_{p_{1}}\left(q+1\right)\delta_{p,p_{1}+1}\delta_{p_{2},p_{1}+2}}{-\sqrt{f_{p_{1}}\left(q\right)f_{p_{2}}\left(q\right)}\eta\left(q\right)-f_{p}\left(q\right)\eta\left(q\right)}\\
 & =\frac{-\eta^{2}\left(q+1\right)}{2\eta\left(q\right)}\frac{f_{p_{2}-1}\left(q+1\right)f_{p_{1}}\left(q+1\right)\delta_{p,p_{1}+1}\delta_{p_{2},p_{1}+2}}{\sqrt{f_{p_{1}}\left(q\right)f_{p_{2}}\left(q\right)}+f_{p_{1}+1}\left(q\right)}
\end{split}
\end{equation}
so
\begin{equation}
\begin{split}\sum_{\beta_{-}}\sum_{\alpha}\frac{\left(\left\langle \beta_{-}\left|H_{1}\right|\alpha\right\rangle \right)^{2}}{E_{\beta_{-}}^{\left(0\right)}-E_{\alpha}^{0}} & =\sum_{p_{1}=0}^{\infty}\sum_{p_{2}=p_{1}+1}^{\infty}\sum_{p=0}^{\infty}M_{1}\\
 & =\frac{-\eta^{2}\left(q+1\right)}{2\eta\left(q\right)}\sum_{p_{1}=0}^{\infty}\sum_{p_{2}=p_{1}+1}^{\infty}\frac{f_{p_{2}-1}\left(q+1\right)f_{p_{1}}\left(q+1\right)\delta_{p_{2},p_{1}+2}}{\sqrt{f_{p_{1}}\left(q\right)f_{p_{2}}\left(q\right)}+f_{p_{1}+1}\left(q\right)}\\
 & =\frac{-\eta^{2}\left(q+1\right)}{2\eta\left(q\right)}\sum_{p_{1}=0}^{\infty}\sum_{p_{3}=p_{1}}^{\infty}\frac{f_{p_{3}}\left(q+1\right)f_{p_{1}}\left(q+1\right)\delta_{p_{3},p_{1}+1}}{\sqrt{f_{p_{1}}\left(q\right)f_{p_{3}+1}\left(q\right)}+f_{p_{1}+1}\left(q\right)}\\
 & =\frac{-\eta^{2}\left(q+1\right)}{2\eta\left(q\right)}\sum_{p_{1}=0}^{\infty}\sum_{p_{2}=p_{1}}^{\infty}\frac{f_{p_{2}}\left(q+1\right)f_{p_{1}}\left(q+1\right)\delta_{p_{2},p_{1}+1}}{\sqrt{f_{p_{1}}\left(q\right)f_{p_{2}+1}\left(q\right)}+\sqrt{f_{p_{1}+1}\left(q\right)f_{p_{2}}\left(q\right)}}\\
 & =\frac{-\eta^{2}\left(q+1\right)}{4\eta\left(q\right)}\sum_{p_{1}=0}^{\infty}\sum_{p_{2}=0}^{\infty}2g\left(p_{1},p_{2}\right)\delta_{p_{2},p_{1}+1}
\end{split}
\end{equation}
here we do a replace that is $p_{2}\rightarrow p_{3}+1$, then let
$p_{3}\rightarrow p_{2}$ .
\begin{equation}
\sum_{\beta_{-}}\sum_{\alpha}\frac{\left(\left\langle \beta_{-}\left|H_{1}\right|\alpha\right\rangle \right)^{2}}{E_{\beta_{-}}^{\left(0\right)}-E_{\alpha}^{0}}=\sum_{p_{1}=0}^{\infty}\sum_{p_{2}=p_{1}+1}^{\infty}M_{1}=\frac{-\eta^{2}\left(q+1\right)}{4\eta\left(q\right)}\sum_{p_{1}=0}^{\infty}\sum_{p_{2}=0}^{\infty}2g\left(p_{1},p_{2}\right)\delta_{p_{2},p_{1}+1}
\end{equation}
Then calculate $\sum_{\beta_{-}}\sum_{\alpha_{+}}\frac{\left(\left\langle \beta_{-}\left|H_{1}\right|\alpha_{+}\right\rangle \right)^{2}}{E_{\beta_{-}}^{\left(0\right)}-E_{\alpha_{+}}^{0}}$.
We do some preparation work firstly, the following four factor form
may be useful 
\begin{equation}
\begin{array}{cc}
A & =\left\langle p_{2}-1,p_{1}+q+1\mid a,b+q\right\rangle =\delta_{a,p_{2}-1}\delta_{b,p_{1}+1}\\
B & =\left\langle p_{2}-1,p_{1}+q+1\mid b,a+q\right\rangle =\delta_{a,p_{1}+1}\delta_{b,p_{2}-1}\\
C & =\left\langle p_{1}-1,p_{2}+q+1\mid a,b+q\right\rangle =\delta_{a,p_{1}-1}\delta_{b,p_{2}+1}\\
D & =\left\langle p_{1}-1,p_{2}+q+1\mid b,a+q\right\rangle =\delta_{a,p_{2}+1}\delta_{b,p_{1}-1}
\end{array}
\end{equation}
and it easy to find that $AD=0,$ $BC=0$ that is 
\begin{equation}
\delta_{a,p_{2}-1}\delta_{b,p_{1}+1}\delta_{a,p_{2}+1}\delta_{b,p_{1}-1}=0
\end{equation}
\begin{equation}
\delta_{a,p_{1}+1}\delta_{b,p_{2}-1}\delta{}_{a,p_{1}-1}\delta_{b,p_{2}+1}=0
\end{equation}
If $p_{1}\neq p_{2}$ ,we have $AC=0,$ $BD=0$ that is
\begin{equation}
\delta_{a,p_{2}-1}\delta_{b,p_{1}+1}\delta_{a,p_{1}-1}\delta_{b,p_{2}+1}=0
\end{equation}
\begin{equation}
\delta_{a,p_{1}+1}\delta_{b,p_{2}-1}\delta_{a,p_{2}+1}\delta_{b,p_{1}-1}=0
\end{equation}
the following factor form may be useful
\begin{equation}
\begin{split} & \left\langle \beta_{-}\right|H_{1}\frac{\left|a,b+q\right\rangle +\left|b,a+q\right\rangle }{\sqrt{2}}\\
= & \frac{\left\langle p_{1},p_{2}+q\right|-\left\langle p_{2},p_{1}+q\right|}{\sqrt{2}}H_{1}\frac{\left|a,b+q\right\rangle +\left|b,a+q\right\rangle }{\sqrt{2}}\\
= & \frac{\eta\left(q+1\right)}{2}\sqrt{f_{p_{2}-1}\left(q+1\right)f_{p_{1}}\left(q+1\right)}\left\langle p_{2}-1,p_{1}+q+1\right|\left[\left|a,b+q\right\rangle +\left|b,a+q\right\rangle \right]\\
- & \frac{\eta\left(q+1\right)}{2}\sqrt{f_{p_{1}-1}\left(q+1\right)f_{p_{2}}\left(q+1\right)}\left\langle p_{1}-1,p_{2}+q+1\right|\left[\left|a,b+q\right\rangle +\left|b,a+q\right\rangle \right]\\
= & \frac{\eta\left(q+1\right)}{2}\sqrt{f_{p_{2}-1}\left(q+1\right)f_{p_{1}}\left(q+1\right)}\left[\delta_{a,p_{2}-1}\delta_{b,p_{1}+1}+\delta_{a,p_{1}+1}\delta_{b,p_{2}-1}\right]\\
- & \frac{\eta\left(q+1\right)}{2}\sqrt{f_{p_{1}-1}\left(q+1\right)f_{p_{2}}\left(q+1\right)}\left[\delta_{a,p_{1}-1}\delta_{b,p_{2}+1}+\delta_{a,p_{2}+1}\delta_{b,p_{1}-1}\right]
\end{split}
\end{equation}
because $p_{1}\neq p_{2}$:
\begin{equation}
\begin{split} & \left[\left\langle \beta_{-}\right|H_{1}\frac{\left|a,b+q\right\rangle +\left|b,a+q\right\rangle }{\sqrt{2}}\right]^{2}\\
= & \frac{\eta^{2}\left(q+1\right)}{4}f_{p_{2}-1}\left(q+1\right)f_{p_{1}}\left(q+1\right)\left[\delta_{a,p_{2}-1}\delta_{b,p_{1}+1}+\delta_{a,p_{1}+1}\delta_{b,p_{2}-1}\right]^{2}\\
+ & \frac{\eta^{2}\left(q+1\right)}{4}f_{p_{1}-1}\left(q+1\right)f_{p_{2}}\left(q+1\right)\left[\delta_{a,p_{1}-1}\delta_{b,p_{2}+1}+\delta_{a,p_{2}+1}\delta_{b,p_{1}-1}\right]^{2}
\end{split}
\end{equation}
if we consider $p_{1}<p_{2}$,and $a<b$,we have 
\begin{equation}
\begin{split} & \left[\left\langle \beta_{-}\right|H_{1}\frac{\left|a,b+q\right\rangle +\left|b,a+q\right\rangle }{\sqrt{2}}\right]^{2}\\
= & \frac{\eta^{2}\left(q+1\right)}{4}f_{p_{2}-1}\left(q+1\right)f_{p_{1}}\left(q+1\right)\delta_{a,p_{2}-1}\delta_{b,p_{1}+1}\delta_{p_{2},p_{1}+1}\\
+ & \frac{\eta^{2}\left(q+1\right)}{4}f_{p_{2}-1}\left(q+1\right)f_{p_{1}}\left(q+1\right)\left[\delta_{a,p_{1}+1}\delta_{b,p_{2}-1}\&\&\left(p_{1}+2<p_{2}\right)\right]\\
+ & \frac{\eta^{2}\left(q+1\right)}{4}f_{p_{1}-1}\left(q+1\right)f_{p_{2}}\left(q+1\right)\left[\delta_{a,p_{1}-1}\delta_{b,p_{2}+1}\&\&\left(p_{1}<p_{2}\right)\right]
\end{split}
\end{equation}
so we have divided the nonezero term of $\frac{\left(\left\langle \beta_{-}\left|H_{1}\right|\alpha_{+}\right\rangle \right)^{2}}{E_{\beta_{-}}^{\left(0\right)}-E_{\alpha_{+}}^{0}}$
into three parts, the first part is 
\begin{equation}
\begin{split}M_{2} & =\frac{\frac{\eta^{2}\left(q+1\right)}{4}f_{p_{2}-1}\left(q+1\right)f_{p_{1}}\left(q+1\right)\delta_{a,p_{2}-1}\delta_{b,p_{1}+1}\delta_{p_{2},p_{1}+1}}{-\sqrt{f_{p_{1}}\left(q\right)f_{p_{2}}\left(q\right)}\eta\left(q\right)-\sqrt{f_{a}\left(q\right)f_{b}\left(q\right)}\eta\left(q\right)}\\
 & =\frac{-\eta^{2}\left(q+1\right)}{4\eta\left(q\right)}\frac{f_{p_{2}-1}\left(q+1\right)f_{p_{1}}\left(q+1\right)\delta_{p_{2},p_{1}+1}}{\sqrt{f_{p_{1}}\left(q\right)f_{p_{2}}\left(q\right)}+\sqrt{f_{p_{1}+1}\left(q\right)f_{p_{2}-1}\left(q\right)}}
\end{split}
\end{equation}
\begin{equation}
\begin{split}\sum_{p_{1}=0}^{\infty}\sum_{p_{2}=p_{1}+1}^{\infty}M_{2} & =\frac{-\eta^{2}\left(q+1\right)}{4\eta\left(q\right)}\sum_{p_{1}=0}^{\infty}\sum_{p_{2}=p_{1}+1}^{\infty}\frac{f_{p_{2}-1}\left(q+1\right)f_{p_{1}}\left(q+1\right)\delta_{p_{2},p_{1}+1}}{\sqrt{f_{p_{1}}\left(q\right)f_{p_{2}}\left(q\right)}+\sqrt{f_{p_{1}+1}\left(q\right)f_{p_{2}-1}\left(q\right)}}\\
 & =\frac{-\eta^{2}\left(q+1\right)}{4\eta\left(q\right)}\sum_{p_{1}=0}^{\infty}\sum_{p_{3}=p_{1}}^{\infty}\frac{f_{p_{3}}\left(q+1\right)f_{p_{1}}\left(q+1\right)\delta_{p_{3},p_{1}}}{\sqrt{f_{p_{1}}\left(q\right)f_{p_{3}+1}\left(q\right)}+\sqrt{f_{p_{1}+1}\left(q\right)f_{p_{3}}\left(q\right)}}\\
 & =\frac{-\eta^{2}\left(q+1\right)}{4\eta\left(q\right)}\sum_{p_{1}=0}^{\infty}\sum_{p_{2}=p_{1}}^{\infty}\frac{f_{p_{2}}\left(q+1\right)f_{p_{1}}\left(q+1\right)\delta_{p_{2},p_{1}}}{\sqrt{f_{p_{1}}\left(q\right)f_{p_{2}+1}\left(q\right)}+\sqrt{f_{p_{1}+1}\left(q\right)f_{p_{2}}\left(q\right)}}\\
 & =\frac{-\eta^{2}\left(q+1\right)}{4\eta\left(q\right)}\sum_{p_{1}=0}^{\infty}\sum_{p_{2}=0}^{\infty}g\left(p_{1},p_{2}\right)\delta_{p_{2},p_{1}}
\end{split}
\end{equation}
so
\begin{equation}
\sum_{p_{1}=0}^{\infty}\sum_{p_{2}=p_{1}+1}^{\infty}M_{2}=\frac{-\eta^{2}\left(q+1\right)}{4\eta\left(q\right)}\sum_{p_{1}=0}^{\infty}\sum_{p_{2}=0}^{\infty}g\left(p_{1},p_{2}\right)\delta_{p_{2},p_{1}}
\end{equation}
 the second part is
\begin{equation}
\begin{split}M_{3} & =\frac{\frac{\eta^{2}\left(q+1\right)}{4}f_{p_{2}-1}\left(q+1\right)f_{p_{1}}\left(q+1\right)\left[\delta_{a,p_{1}+1}\delta_{b,p_{2}-1}\&\&\left(p_{1}+2<p_{2}\right)\right]}{-\sqrt{f_{p_{1}}\left(q\right)f_{p_{2}}\left(q\right)}\eta\left(q\right)-\sqrt{f_{a}\left(q\right)f_{b}\left(q\right)}\eta\left(q\right)}\\
 & =\frac{-\eta^{2}\left(q+1\right)}{4\eta\left(q\right)}\frac{f_{p_{2}-1}\left(q+1\right)f_{p_{1}}\left(q+1\right)\left[p_{1}+2<p_{2}\right]}{\sqrt{f_{p_{1}}\left(q\right)f_{p_{2}}\left(q\right)}+\sqrt{f_{p_{1}+1}\left(q\right)f_{p_{2}-1}\left(q\right)}}
\end{split}
\end{equation}
\begin{equation}
\begin{split}\sum_{p_{1}=0}^{\infty}\sum_{p_{2}=p_{1}+1}^{\infty}M_{3} & =\frac{-\eta^{2}\left(q+1\right)}{4\eta\left(q\right)}\sum_{p_{1}=0}^{\infty}\sum_{p_{2}=p_{1}+1}^{\infty}\frac{f_{p_{2}-1}\left(q+1\right)f_{p_{1}}\left(q+1\right)\left[p_{1}+2<p_{2}\right]}{\sqrt{f_{p_{1}}\left(q\right)f_{p_{2}}\left(q\right)}+\sqrt{f_{p_{1}+1}\left(q\right)f_{p_{2}-1}\left(q\right)}}\\
 & =\frac{-\eta^{2}\left(q+1\right)}{4\eta\left(q\right)}\sum_{p_{1}=0}^{\infty}\sum_{p_{3}=p_{1}}^{\infty}\frac{f_{p_{3}}\left(q+1\right)f_{p_{1}}\left(q+1\right)\left[p_{1}+1<p_{3}\right]}{\sqrt{f_{p_{1}}\left(q\right)f_{p_{3}+1}\left(q\right)}+\sqrt{f_{p_{1}+1}\left(q\right)f_{p_{3}}\left(q\right)}}\\
 & =\frac{-\eta^{2}\left(q+1\right)}{4\eta\left(q\right)}\sum_{p_{1}=0}^{\infty}\sum_{p_{2}=p_{1}}^{\infty}\frac{f_{p_{2}}\left(q+1\right)f_{p_{1}}\left(q+1\right)\left[p_{1}+1<p_{2}\right]}{\sqrt{f_{p_{1}}\left(q\right)f_{p_{2}+1}\left(q\right)}+\sqrt{f_{p_{1}+1}\left(q\right)f_{p_{2}}\left(q\right)}}\\
 & =\frac{-\eta^{2}\left(q+1\right)}{4\eta\left(q\right)}\sum_{p_{1}=0}^{\infty}\sum_{p_{2}=p_{1}+2}^{\infty}\frac{f_{p_{2}}\left(q+1\right)f_{p_{1}}\left(q+1\right)}{\sqrt{f_{p_{1}}\left(q\right)f_{p_{2}+1}\left(q\right)}+\sqrt{f_{p_{1}+1}\left(q\right)f_{p_{2}}\left(q\right)}}\\
 & =\frac{-\eta^{2}\left(q+1\right)}{4\eta\left(q\right)}\sum_{p_{1}=0}^{\infty}\sum_{p_{2}=p_{1}+2}^{\infty}g\left(p_{1},p_{2}\right)
\end{split}
\end{equation}
so
\begin{equation}
\sum_{p_{1}=0}^{\infty}\sum_{p_{2}=p_{1}+1}^{\infty}M_{3}=\frac{-\eta^{2}\left(q+1\right)}{4\eta\left(q\right)}\sum_{p_{1}=0}^{\infty}\sum_{p_{2}=p_{1}+2}^{\infty}g\left(p_{1},p_{2}\right)
\end{equation}
the third part is 
\begin{equation}
\begin{split}M_{4} & =\frac{\frac{\eta^{2}\left(q+1\right)}{4}f_{p_{1}-1}\left(q+1\right)f_{p_{2}}\left(q+1\right)\left[\delta_{a,p_{1}-1}\delta_{b,p_{2}+1}\&\&\left(p_{1}<p_{2}\right)\right]}{-\sqrt{f_{p_{1}}\left(q\right)f_{p_{2}}\left(q\right)}\eta\left(q\right)-\sqrt{f_{a}\left(q\right)f_{b}\left(q\right)}\eta\left(q\right)}\\
 & =\frac{-\eta^{2}\left(q+1\right)}{4\eta\left(q\right)}\frac{f_{p_{1}-1}\left(q+1\right)f_{p_{2}}\left(q+1\right)\left[p_{1}<p_{2}\right]}{\sqrt{f_{p_{1}}\left(q\right)f_{p_{2}}\left(q\right)}+\sqrt{f_{p_{1}-1}\left(q\right)f_{p_{2}+1}\left(q\right)}}
\end{split}
\end{equation}
\begin{equation}
\begin{split}\sum_{p_{1}=1}^{\infty}\sum_{p_{2}=p_{1}+1}^{\infty}M_{4} & =\frac{-\eta^{2}\left(q+1\right)}{4\eta\left(q\right)}\sum_{p_{1}=1}^{\infty}\sum_{p_{2}=p_{1}+1}^{\infty}\frac{f_{p_{1}-1}\left(q+1\right)f_{p_{2}}\left(q+1\right)\left[p_{1}<p_{2}\right]}{\sqrt{f_{p_{1}}\left(q\right)f_{p_{2}}\left(q\right)}+\sqrt{f_{p_{1}-1}\left(q\right)f_{p_{2}+1}\left(q\right)}}\\
 & =\frac{-\eta^{2}\left(q+1\right)}{4\eta\left(q\right)}\sum_{p_{3}=0}^{\infty}\sum_{p_{2}=p_{3}+2}^{\infty}\frac{f_{p_{3}}\left(q+1\right)f_{p_{2}}\left(q+1\right)\left[p_{3}+1<p_{2}\right]}{\sqrt{f_{p_{3}+1}\left(q\right)f_{p_{2}}\left(q\right)}+\sqrt{f_{p_{3}}\left(q\right)f_{p_{2}+1}\left(q\right)}}\\
 & =\frac{-\eta^{2}\left(q+1\right)}{4\eta\left(q\right)}\sum_{p_{1}=0}^{\infty}\sum_{p_{2}=p_{1}+2}^{\infty}\frac{f_{p_{1}}\left(q+1\right)f_{p_{2}}\left(q+1\right)}{\sqrt{f_{p_{1}+1}\left(q\right)f_{p_{2}}\left(q\right)}+\sqrt{f_{p_{1}}\left(q\right)f_{p_{2}+1}\left(q\right)}}\\
 & =\frac{-\eta^{2}\left(q+1\right)}{4\eta\left(q\right)}\sum_{p_{1}=0}^{\infty}\sum_{p_{2}=p_{1}+2}^{\infty}g\left(p_{1},p_{2}\right)
\end{split}
\end{equation}
\begin{equation}
\sum_{p_{1}=1}^{\infty}\sum_{p_{2}=p_{1}+1}^{\infty}M_{4}=\frac{-\eta^{2}\left(q+1\right)}{4\eta\left(q\right)}\sum_{p_{1}=0}^{\infty}\sum_{p_{2}=p_{1}+2}^{\infty}g\left(p_{1},p_{2}\right)
\end{equation}
so 
\begin{equation}
\sum_{\alpha_{+}}\frac{\left(\left\langle \beta_{-}\left|H_{1}\right|\alpha_{+}\right\rangle \right)^{2}}{E_{\beta_{-}}^{\left(0\right)}-E_{\alpha_{+}}^{0}}=M_{2}+M_{3}+M_{4}
\end{equation}
and
\begin{equation}
\begin{split}\sum_{\beta_{-}}\sum_{\alpha_{+}}\frac{\left(\left\langle \beta_{-}\left|H_{1}\right|\alpha_{+}\right\rangle \right)^{2}}{E_{\beta_{-}}^{\left(0\right)}-E_{\alpha_{+}}^{0}} & =\sum_{p_{1}=0}^{\infty}\sum_{p_{2}=p_{1}+1}^{\infty}M_{2}+\sum_{p_{1}=0}^{\infty}\sum_{p_{2}=p_{1}+1}^{\infty}M_{3}+\sum_{p_{1}=1}^{\infty}\sum_{p_{2}=p_{1}+1}^{\infty}M_{4}\\
 & =\frac{-\eta^{2}\left(q+1\right)}{4\eta\left(q\right)}\left[\sum_{p_{1}=0}^{\infty}\sum_{p_{2}=0}^{\infty}g\left(p_{1},p_{2}\right)\delta_{p_{2},p_{1}}+\sum_{p_{1}=0}^{\infty}\sum_{p_{2}=p_{1}+2}^{\infty}2g\left(p_{1},p_{2}\right)\right]
\end{split}
\end{equation}
so we can get a simple expression of $\sum_{\beta_{-}}E_{\beta_{-}}^{\left(2\right)}$

\begin{equation}
\begin{split} & \sum_{\beta_{-}}E_{\beta_{-}}^{\left(2\right)}\\
= & \sum_{\beta_{-}}\sum_{\alpha_{+}}\frac{\left(\left\langle \beta_{-}\left|H_{1}\right|\alpha_{+}\right\rangle \right)^{2}}{E_{\beta_{-}}^{\left(0\right)}-E_{\alpha_{+}}^{0}}+\sum_{\beta_{-}}\sum_{\alpha}\frac{\left(\left\langle \beta_{-}\left|H_{1}\right|\alpha\right\rangle \right)^{2}}{E_{\beta_{-}}^{\left(0\right)}-E_{\alpha}^{0}}\\
= & \frac{-\eta^{2}\left(q+1\right)}{4\eta\left(q\right)}\left[\sum_{p_{1}=0}^{\infty}\sum_{p_{2}=0}^{\infty}g\left(p_{1},p_{2}\right)\delta_{p_{2},p_{1}}+\sum_{p_{1}=0}^{\infty}\sum_{p_{2}=p_{1}+2}^{\infty}2g\left(p_{1},p_{2}\right)+\sum_{p_{1}=0}^{\infty}\sum_{p_{2}=0}^{\infty}2g\left(p_{1},p_{2}\right)\delta_{p_{2},p_{1}+1}\right]\\
= & \frac{-\eta^{2}\left(q+1\right)}{4\eta\left(q\right)}\left[\sum_{p_{1}=0}^{\infty}g\left(p_{1},p_{1}\right)+\sum_{p_{1}=0}^{\infty}\sum_{p_{1}<p_{2}}^{\infty}2g\left(p_{1},p_{2}\right)\right]\\
= & \frac{-\eta^{2}\left(q+1\right)}{4\eta\left(q\right)}\left[\sum_{p_{1}=0}^{\infty}\sum_{p_{2}=0}^{\infty}g\left(p_{1},p_{2}\right)\delta_{p_{2},p_{1}}+\sum_{p_{1}=0}^{\infty}\sum_{p_{1}<p_{2}}^{\infty}g\left(p_{1},p_{2}\right)+\sum_{p_{2}=0}^{\infty}\sum_{p_{2}<p_{1}}^{\infty}g\left(p_{1},p_{2}\right)\right]\\
= & \frac{-\eta^{2}\left(q+1\right)}{4\eta\left(q\right)}\sum_{p_{1}=0}^{\infty}\sum_{p_{2}=0}^{\infty}g\left(p_{1},p_{2}\right)
\end{split}
\end{equation}
\begin{equation}
\sum_{\beta_{-}}E_{\beta_{-}}^{\left(2\right)}=\frac{-\eta^{2}\left(q+1\right)}{4\eta\left(q\right)}\sum_{p_{1}=0}^{\infty}\sum_{p_{2}=0}^{\infty}g\left(p_{1},p_{2}\right)
\end{equation}
If we consider second order, maybe we should calculate the contribution
of the perturbation Hamiltonian $H_{2}$ when $s=q+2$ .
\[
H_{2}=\sum_{i=0}^{\infty}\sum_{,j=0}^{\infty}\sqrt{f_{i}\left(q+2\right)f_{j}\left(q+2\right)}\eta\left(q+2\right)\left|i,j+q+2\right\rangle \left\langle j,i+q+2\right|
\]
and the following formulas may be useful
\begin{equation}
\begin{split} & \langle\beta_{-}\left|i,j+q+2\right\rangle \left\langle j,i+q+2\right|\\
= & \frac{\left\langle p_{1},p_{2}+q\right|-\left\langle p_{2},p_{1}+q\right|}{\sqrt{2}}\left|i,j+q+2\right\rangle \left\langle j,i+q+2\right|\\
= & \frac{\left\langle p_{2}-2,p_{1}+q+2\right|}{\sqrt{2}}-\frac{\left\langle p_{1}-2,p_{2}+q+2\right|}{\sqrt{2}}
\end{split}
\end{equation}
 
\begin{equation}
\begin{split} & \left\langle \beta_{-}\right|H_{2}\\
 & \frac{\left\langle p_{1},p_{2}+q\right|-\left\langle p_{2},p_{1}+q\right|}{\sqrt{2}}H_{2}\\
= & P\left(q+2\right)\sqrt{f_{p_{2}-2}\left(q+2\right)f_{p_{1}}\left(q+2\right)}\frac{\left\langle p_{2}-2,p_{1}+q+2\right|}{\sqrt{2}}\\
- & P\left(q+2\right)\sqrt{f_{p_{1}-2}\left(q+2\right)f_{p_{2}}\left(q+2\right)}\frac{\left\langle p_{1}-2,p_{2}+q+2\right|}{\sqrt{2}}
\end{split}
\end{equation}
we calculate$\left\langle \beta_{-}\right|H_{2}\left|\beta_{-}\right\rangle $:

\begin{equation}
\begin{split} & \left\langle \beta_{-}\right|H_{2}\left|\beta_{-}\right\rangle \\
= & \eta\left(q+2\right)\sqrt{f_{p_{2}-2}\left(q+2\right)f_{p_{1}}\left(q+2\right)}\frac{\left\langle p_{2}-2,p_{1}+q+2\right|}{\sqrt{2}}\frac{\left|p_{1},p_{2}+q\right\rangle -\left|p_{2},p_{1}+q\right\rangle }{\sqrt{2}}\\
- & \eta\left(q+2\right)\sqrt{f_{p_{1}-2}\left(q+2\right)f_{p_{2}}\left(q+2\right)}\frac{\left\langle p_{1}-2,p_{2}+q+2\right|}{\sqrt{2}}\frac{\left|p_{1},p_{2}+q\right\rangle -\left|p_{2},p_{1}+q\right\rangle }{\sqrt{2}}\\
= & \frac{\eta\left(q+2\right)}{2}\left(\sqrt{f_{p_{2}-2}\left(q+2\right)f_{p_{1}}\left(q+2\right)}\delta_{p_{2},p_{1}+2}+\sqrt{f_{p_{1}-2}\left(q+2\right)f_{p_{2}}\left(q+2\right)}\delta_{p_{1},p_{2}+2}\right)\\
= & \frac{\eta\left(q+2\right)}{2}\left(f_{p_{1}}\left(q+2\right)\delta_{p_{2},p_{1}+2}+f_{p_{2}}\left(q+2\right)\delta_{p_{1},p_{2}+2}\right)
\end{split}
\end{equation}
 
\[
\begin{split}M_{5} & =-\sum_{k_{-}}\left\langle \beta_{-}\right|H_{2}\left|\beta_{-}\right\rangle \\
 & =-\frac{1}{2}\sum_{p_{1}\neq p_{2}}^{\infty}\frac{\eta\left(q+2\right)}{2}\left(f_{p_{1}}\left(q+2\right)\delta_{p_{2},p_{1}+2}+f_{p_{2}}\left(q+2\right)\delta_{p_{1},p_{2}+2}\right)\\
 & =-\frac{\eta\left(q+2\right)}{4}\left(\sum_{p_{1}\neq p_{2}}^{\infty}f_{p_{1}}\left(q+2\right)\delta_{p_{2},p_{1}+2}+\sum_{p_{1}\neq p_{2}}^{\infty}f_{p_{2}}\left(q+2\right)\delta_{p_{1},p_{2}+2}\right)\\
 & =-\frac{\eta\left(q+2\right)}{2}
\end{split}
\]
 If we just consider zero order, first order and second order ,
\begin{equation}
E_{N}^{\mu}\left(q\right)=\ln\left[1+2\left(\sum_{\beta_{-}}\left(-E_{\beta_{-}}^{\left(0\right)}-E_{\beta_{-}}^{\left(1\right)}-E_{\beta_{-}}^{\left(2\right)}\right)+M_{5}\right)\right]
\end{equation}
with
\[
\begin{split}\sum_{\beta_{-}}\left(-E_{\beta_{-}}^{\left(0\right)}\right) & =\frac{1}{2}\left(\sum_{p=0}^{\infty}\sqrt{f_{p}\left(q\right)}\right)^{2}\eta\left(q\right)-\frac{1}{2}\eta\left(q\right)\\
\sum_{\beta_{-}}\left(-E_{\beta_{-}}^{\left(1\right)}\right) & =-\frac{\eta\left(q+1\right)}{2}\\
\sum_{\beta_{-}}\left(-E_{\beta_{-}}^{\left(2\right)}\right) & =\frac{\eta^{2}\left(q+1\right)}{4\eta\left(q\right)}\sum_{p_{1}=0}^{\infty}\sum_{p_{2}=0}^{\infty}g\left(p_{1},p_{2}\right)\\
M_{5} & =-\frac{\eta\left(q+2\right)}{2}
\end{split}
\]
so
\begin{equation}
\begin{split} & E_{N}^{\mu}\left(q\right)\\
= & \ln[1+\left(\sum_{p=0}^{\infty}\sqrt{f_{p}\left(q\right)}\right)^{2}\eta\left(q\right)-\eta\left(q\right)-\eta\left(q+1\right)\\
 & -\eta\left(q+2\right)+\frac{\eta^{2}\left(q+1\right)}{2\eta\left(q\right)}\sum_{p_{1}=0}^{\infty}\sum_{p_{2}=0}^{\infty}g\left(p_{1},p_{2}\right)]
\end{split}
\end{equation}
and we have calculated $\eta\left(q\right)$ and $\eta\left(q+1\right)$

\begin{equation}
\eta\left(q\right)=\left(\frac{1+N_{m}\mu}{1+N_{m}}\right)^{1+q}=\left(1+\frac{N_{m}\left(\mu-1\right)}{1+N_{m}}\right)^{1+q}=\left(1-\varepsilon\right)^{1+q}
\end{equation}
\begin{equation}
\eta\left(q+1\right)=\frac{\left(q+1\right)\left(1-\mu\right)N_{m}}{1+N_{m}}\left(\frac{1+N_{m}\mu}{1+N_{m}}\right)^{1+q}=\left(q+1\right)\varepsilon\left(1-\varepsilon\right)^{1+q}
\end{equation}
\begin{equation}
\eta\left(q+2\right)=\frac{\left(q+1\right)\left(q+2\right)\left(1-\mu\right)^{2}N_{m}^{2}}{2\left(1+N_{m}\right)^{2}}\left(\frac{1+N_{m}\mu}{1+N_{m}}\right)^{1+q}=\frac{\left(q+1\right)\left(q+2\right)}{2}\varepsilon^{2}\left(1-\varepsilon\right)^{1+q}
\end{equation}
\begin{equation}
\frac{\eta^{2}\left(q+1\right)}{\eta^{2}\left(q\right)}=\frac{\left(q+1\right)^{2}\varepsilon^{2}\left(1-\varepsilon\right)^{2+2q}}{\left(1-\varepsilon\right)^{2+2q}}=\left(q+1\right)^{2}\varepsilon^{2}
\end{equation}
here $\varepsilon=\frac{\left(1-\mu\right)N_{m}}{1+N_{m}}$ , and
use the approximation formula $\left(1+x\right)^{n}\approx1+nx+\frac{n\left(n-1\right)}{2}x^{2}$.
\begin{equation}
\left(1-\varepsilon\right)^{1+q}=1-\left(q+1\right)\varepsilon+\frac{q\left(q+1\right)}{2}\varepsilon^{2}
\end{equation}
\begin{equation}
\eta\left(q\right)=\left(1-\varepsilon\right)^{1+q}\approx1-\left(q+1\right)\varepsilon+\frac{q\left(q+1\right)}{2}\varepsilon^{2}
\end{equation}
\begin{equation}
\eta\left(q+1\right)=\left(q+1\right)\varepsilon\left(1-\varepsilon\right)^{1+q}\approx\left(q+1\right)\varepsilon-\left(q+1\right)^{2}\varepsilon^{2}
\end{equation}
\begin{equation}
\eta\left(q+2\right)\approx\frac{\left(q+1\right)\left(q+2\right)}{2}\varepsilon^{2}
\end{equation}
so
\[
1-\eta\left(q\right)-\eta\left(q+1\right)-\eta\left(q+2\right)=0
\]
and
\begin{equation}
E_{N}^{\mu}\left(q\right)=\ln\left[\left(\sum_{p=0}^{\infty}\sqrt{f_{p}\left(q\right)}\right)^{2}\eta\left(q\right)+\frac{\eta^{2}\left(q+1\right)}{2\eta\left(q\right)}\sum_{p_{1}=0}^{\infty}\sum_{p_{2}=0}^{\infty}g\left(p_{1},p_{2}\right)\right]
\end{equation}
\begin{equation}
E_{N}^{\mu}\left(q\right)=\ln\left[\left(\sum_{p=0}^{\infty}\sqrt{f_{p}\left(q\right)}\right)^{2}\eta\left(q\right)\right]+\ln\left[1+\frac{\eta^{2}\left(q+1\right)}{2\eta^{2}\left(q\right)}\left(\frac{\sum_{p_{1}=0}^{\infty}\sum_{p_{2}=0}^{\infty}g\left(p_{1},p_{2}\right)}{\left(\sum_{p=0}^{\infty}\sqrt{f_{p}\left(q\right)}\right)^{2}}\right)\right]
\end{equation}
Define 
\[
\Omega=\frac{\sum_{p_{1}=0}^{\infty}\sum_{p_{2}=0}^{\infty}g\left(p_{1},p_{2}\right)}{\left(\sum_{p=0}^{\infty}\sqrt{f_{p}\left(q\right)}\right)^{2}}
\]

\begin{equation}
E_{N}^{\mu}\left(q\right)=2\ln\sum_{p=0}^{\infty}\sqrt{f_{p}\left(q\right)}+\ln\eta\left(q\right)+\ln\left(1+\frac{\eta^{2}\left(q+1\right)}{2\eta^{2}\left(q\right)}\Omega\right)
\end{equation}
make a further approximation
\begin{equation}
\ln\eta\left(q\right)=\ln\left(1-\left((q+1\right)\varepsilon+\frac{q\left(q+1\right)}{2}\varepsilon^{2}\right)\approx-\left(q+1\right)\varepsilon-\frac{\left(q+1\right)}{2}\varepsilon^{2}
\end{equation}
\begin{equation}
\ln\left(1+\frac{\eta^{2}\left(q+1\right)}{2\eta^{2}\left(q\right)}\Omega\right)\approx\frac{\eta^{2}\left(q+1\right)}{2\eta^{2}\left(q\right)}\Omega=\frac{\left(q+1\right)^{2}\varepsilon^{2}}{2}\Omega
\end{equation}
\begin{equation}
E_{N}^{\mu}\left(q\right)=2\ln\sum_{p=0}^{\infty}\sqrt{f_{p}\left(q\right)}-\left(q+1\right)\varepsilon-\frac{\left(q+1\right)}{2}\varepsilon^{2}+\frac{\left(q+1\right)^{2}\varepsilon^{2}}{2}\Omega
\end{equation}
also $E_{N}\left(q\right)=2\ln\sum_{p=0}^{\infty}\sqrt{f_{p}(q)}$
\begin{equation}
E_{N}^{\mu}\left(q\right)=E_{N}\left(q\right)-\left(q+1\right)\varepsilon-\frac{\left(q+1\right)}{2}\varepsilon^{2}+\frac{\left(q+1\right)^{2}\varepsilon^{2}}{2}\Omega
\end{equation}
\begin{equation}
E_{N}^{\mu}\left(q\right)=E_{N}\left(q\right)+\left(q+1\right)\varepsilon\left[\frac{\Omega q+\Omega-1}{2}\varepsilon-1\right]
\end{equation}
 In the Gaussian approximation and under large $q$ we have 
\begin{equation}
\Omega\approx\frac{1}{2}\sqrt{\frac{1+\zeta q}{\zeta+\zeta q}}\approx\frac{1}{2}
\end{equation}
Then we get 
\begin{equation}
E_{N}^{\mu}\left(q\right)=E_{N}\left(q\right)-\left(q+1\right)\varepsilon+\frac{q^{2}-1}{4}\varepsilon^{2}
\end{equation}

\subsection*{Supplement[I]: On-off Detection Measure}

an on-off detection measure is given by 
\begin{equation}
\varPi^{off}=\left|0\right\rangle _{q}\left\langle 0\right|
\end{equation}

\begin{equation}
\varPi^{on}=I-\left|0\right\rangle _{q}\left\langle 0\right|=\sum_{k=1}^{\infty}\left|k\right\rangle _{q}\left\langle k\right|
\end{equation}
and the state after measure is given by 
\begin{equation}
\begin{array}{ccc}
\rho_{off}^{\prime}=Tr_{q}\left[\varPi^{off}\rho_{i}\right] & , & \rho_{off}=\frac{\rho_{off}^{\prime}}{Tr\left[\rho_{off}^{'}\right]}\end{array}
\end{equation}
\begin{equation}
\begin{array}{ccc}
\rho_{on}^{\prime}=Tr_{q}\left[\varPi^{on}\rho_{i}\right] & , & \rho_{on}=\frac{\rho_{on}^{\prime}}{Tr\left[\rho_{on}^{\prime}\right]}\end{array}
\end{equation}
 we first calculate the off detection measure,

\begin{equation}
\begin{array}{ccc}
\rho_{off}^{\prime}=Tr_{q}\left[\varPi^{off}\rho_{i}\right] & , & \rho_{off}=\frac{\rho_{off}^{\prime}}{Tr\left[\rho_{off}^{'}\right]}\end{array}
\end{equation}
 the trace is calculated by the following
\begin{equation}
Tr_{q}\left[\varPi^{off}\rho_{i}\right]=\sum_{q_{3}=0}^{\infty}\left\langle q_{3}\right|\varPi^{off}\rho_{i}\left|q_{3}\right\rangle 
\end{equation}
so 
\begin{equation}
\rho_{off}^{\prime}=\sum_{p_{1}=0}^{\infty}\sum_{p_{2}=0}^{\infty}\sqrt{f_{p_{1}}\left(0\right)}\sqrt{f_{p_{2}}\left(0\right)}P_{0}\left|p_{1},p_{1}\right\rangle \left\langle p_{2},p_{2}\right|
\end{equation}
and
\begin{equation}
\rho_{off}^{\prime}=\left|\psi^{\prime}\right\rangle {}_{off}\left\langle \psi^{\prime}\right|
\end{equation}
with 
\begin{equation}
\left|\psi^{\prime}\right\rangle {}_{off}=\sum_{p=0}^{\infty}\sqrt{f_{p}\left(0\right)P_{0}}\left|p,p\right\rangle 
\end{equation}
the $\mid\psi^{\prime}\rangle_{off}$ is not a normalized state. So
we normalize the $\left|\psi^{\prime}\right\rangle {}_{off}$ , and
the normalization factor of $\left|\psi^{\prime}\right\rangle {}_{off}$
is 

\begin{equation}
_{off}\left\langle \psi^{\prime}\mid\psi^{\prime}\right\rangle _{off}=\sum_{p=0}^{\infty}\left[\sqrt{f_{p}\left(0\right)P_{0}}\right]^{2}=P_{0}=\frac{1}{1+N_{m}}
\end{equation}
and 
\begin{equation}
Tr\left[\rho_{off}^{\prime}\right]=P_{0}
\end{equation}
so the state after off detection measure is given by 
\begin{equation}
\rho_{off}=\frac{\rho_{off}^{\prime}}{Tr\left[\rho_{off}^{\prime}\right]}=\sum_{p_{1}=0}^{\infty}\sum_{p_{2}=0}^{\infty}\sqrt{f_{p_{1}}\left(0\right)}\sqrt{f_{p_{2}}\left(0\right)}\left|p_{1},p_{1}\right\rangle \left\langle p_{2},p_{2}\right|
\end{equation}
then normalize $\left|\psi^{\prime}\right\rangle {}_{off}$ we get
\begin{equation}
\left|\psi\right\rangle {}_{off}=\frac{1}{\sqrt{P_{0}}}\sum_{p=0}^{\infty}\sqrt{f_{p}\left(0\right)P_{0}}\left|p,p\right\rangle =\sum_{p=0}^{\infty}\sqrt{f_{p}\left(0\right)}\left|p,p\right\rangle =\left|\Psi_{0}\right\rangle 
\end{equation}
\[
\rho_{off}=\left|\psi\right\rangle {}_{off}\left\langle \psi\right|=\left|\Psi_{0}\right\rangle \left\langle \Psi_{0}\right|
\]
with 
\begin{equation}
\left|\Psi_{q}\right\rangle =\sum_{p=0}^{\infty}\sqrt{f_{p}\left(q\right)}\left|p,p+q\right\rangle 
\end{equation}
and it's entanglement is
\begin{equation}
En_{off}=2\ln\sum_{p=0}^{\infty}\sqrt{f_{p}\left(0\right)}=\ln\frac{1+\sqrt{\zeta}}{1-\sqrt{\zeta}}
\end{equation}
we now calculate the on detection measure,

\begin{equation}
\begin{array}{ccc}
\rho_{on}^{\prime}=Tr_{q}\left[\varPi^{on}\rho_{i}\right] & , & \rho_{on}=\frac{\rho_{on}^{\prime}}{Tr\left[\rho_{on}^{\prime}\right]}\end{array}
\end{equation}
\begin{equation}
Tr_{q}\left[\varPi^{on}\rho_{i}\right]=\sum_{q_{3}=0}^{\infty}\left\langle q_{3}\right|\varPi^{on}\rho_{i}\left|q_{3}\right\rangle 
\end{equation}
so we get 
\begin{equation}
\begin{split}\rho_{on}^{\prime}= & \sum_{k=1}^{\infty}\sum_{p_{1}=0}^{\infty}\sum_{p_{2}=0}^{\infty}\sqrt{f_{p_{1}}\left(k\right)f_{p_{2}}\left(k\right)}P_{k}\\
 &\times \left|p_{1},p_{1}+k\right\rangle \left\langle p_{2},p_{2}+k\right|
\end{split}
\end{equation}
 and $\left|\Psi_{q}\right\rangle =\sum_{p=0}^{\infty}\sqrt{f_{p}\left(q\right)}\left|p,p+q\right\rangle $
\[
\left|\Psi_{k}\right\rangle \left\langle \Psi_{k}\right|=\sum_{p_{1}=0}^{\infty}\sum_{p_{2}=0}^{\infty}\sqrt{f_{p_{1}}\left(k\right)f_{p_{2}}\left(k\right)}\left|p_{1},p_{1}+k\right\rangle \left\langle p_{2},p_{2}+k\right|
\]
so we have 
\begin{equation}
\rho_{on}^{\prime}=\sum_{k=1}^{\infty}P_{k}\left|\Psi_{k}\right\rangle \left\langle \Psi_{k}\right|
\end{equation}
so 
\begin{equation}
Tr\left[\rho_{on}^{\prime}\right]=\sum_{k=1}^{\infty}P_{k}=1-P_{0}=\frac{N_{m}}{1+N_{m}}
\end{equation}
then the normalized $\rho_{on}$ is 
\begin{equation}
\begin{split}\rho_{on} & =\frac{\rho_{on}^{\prime}}{Tr\left[\rho_{on}^{\prime}\right]}\\
 & =\frac{1+N_{m}}{N_{m}}\sum_{k=1}^{\infty}\sum_{p_{1}=0}^{\infty}\sum_{p_{2}=0}^{\infty}\sqrt{f_{p_{1}}\left(k\right)f_{p_{2}}\left(k\right)}P_{k}\\
 &\times \left|p_{1},p_{1}+k\right\rangle \left\langle p_{2},p_{2}+k\right|
\end{split}
\end{equation}
and
\begin{equation}
\rho_{on}=\frac{1+N_{m}}{N_{m}}\sum_{k=1}^{\infty}P_{k}\left|\Psi_{k}\right\rangle \left\langle \Psi_{k}\right|
\end{equation}
and the next problem is how to calculate the entanglement of $\rho_{on}$.
In perfect measurement we have the average entanglement defined by
\begin{equation}
\overline{E}_{N}=\sum_{q}P_{q}E_{N}\left(q\right)
\end{equation}
then we can define the on average entanglement 
\begin{equation}
\overline{E}_{N}^{on}=\frac{1+N_{m}}{N_{m}}\sum_{k=1}^{\propto}P_{k}E_{N}\left(k\right)
\end{equation}
also we get 
\begin{equation}
\overline{E}_{N}^{on}=\frac{1+N_{m}}{N_{m}}\overline{E}_{N}-\frac{1}{N_{m}}\ln\frac{\sqrt{8\pi\zeta}}{1-\zeta}
\end{equation}

\subsection*{Supplement[J]:  Derivation of the Entanglement's Equation }

The entanglement is give by this equation
\begin{equation}
E_{N}\left(\rho\right)=\ln\left\Vert \rho^{\mathbf{T}}\right\Vert _{1}
\end{equation}
for a simple density matrix the partial transpose is 
\begin{equation}
\rho=\sum_{n,m}C_{n}C_{m}\left|n_{A},n_{B}\right\rangle \left\langle m_{A},m_{B}\right|
\end{equation}
\[
\rho^{\mathbf{PT_{A}}}=\sum_{n,m}C_{n}C_{m}\left|m_{A},n_{B}\right\rangle \left\langle n_{A},m_{B}\right|
\]
or 
\begin{equation}
\rho^{\mathbf{PT_{B}}}=\sum_{n,m}C_{n}C_{m}\left|n_{A},m_{B}\right\rangle \left\langle m_{A},n_{B}\right|
\end{equation}
so 
\begin{equation}
\rho^{\mathbf{PT_{A}}}=\left(\rho^{\mathbf{PT_{B}}}\right)^{\mathbf{T}}
\end{equation}
then we just write $\rho^{\mathbf{T}}=\rho^{\mathbf{PT_{A}}}$. The
trace form is 
\begin{equation}
\left\Vert \rho\right\Vert _{1}=tr\sqrt{\rho^{+}\rho}=\sum_{i}\left|\lambda_{i}\right|
\end{equation}
\begin{equation}
E_{N}\left(\rho\right)=ln\left[\sum_{i}\left|\lambda_{i}\right|\right]=\ln\left[1+2N\left(\rho^{\mathbf{T}}\right)\right]
\end{equation}
where $\left|\lambda_{i}\right|$ is the absolute eigenvalues of $\rho^{\mathbf{T}}$
and $N\left(\rho^{\mathbf{T}}\right)$ is the sum of the absolute
value of the negative eigenvalues of $\rho^{\mathbf{T}}$ , $N\left(\rho^{\mathbf{T}}\right)=\sum_{\lambda_{i}<0}\left|\lambda_{i}\right|$,the
relationship between $\sum_{i}\left|\lambda_{i}\right|$and $N\left(\rho^{\mathbf{T}}\right)$
are given by 
\begin{equation}
\begin{split}\sum_{i}\left|\lambda_{i}\right| & =\sum_{\lambda_{i}\geq0}\left|\lambda_{i}\right|+\sum_{\lambda_{i}<0}\left|\lambda_{i}\right|\\
 & =\sum_{\lambda_{i}\geq0}\lambda_{i}-\sum_{\lambda_{i}<0}\lambda_{i}\\
 & =\sum_{\lambda_{i}\geq0}\lambda_{i}+\sum_{\lambda_{i}<0}\lambda_{i}-\sum_{\lambda_{i}<0}\lambda_{i}-\sum_{\lambda_{i}<0}\lambda_{i}\\
 & =1-2\sum_{\lambda_{i}<0}\lambda_{i}\\
 & =1+2N\left(\rho^{\mathbf{T}}\right)
\end{split}
\end{equation}
Consider a general two mode entangled state
\begin{equation}
\left|\psi\right\rangle =\sum_{n}C_{n}\left|n,n\right\rangle 
\end{equation}
the normalization condition is $\sum_{n}C_{n}^{2}=1$ 
\begin{equation}
\rho=\left|\psi\right\rangle \left\langle \psi\right|=\sum_{n,m}C_{n}C_{m}^{*}\left|n,n\right\rangle \left\langle m,m\right|
\end{equation}
and the partial transpose is 
\begin{equation}
\rho^{\mathbf{T}}=\sum_{n,m}C_{n}C_{m}^{*}\left|n,m\right\rangle \left\langle m,n\right|
\end{equation}
to find the negative eigenvalues of $\rho^{\mathbf{T}}$,we rewrite
$\rho^{\mathbf{T}}$ in this way

\[
2\rho^{\mathbf{T}}=\sum_{n,m}C_{n}C_{m}^{*}\left|n,m\right\rangle \left\langle m,n\right|+\sum_{n,m}C_{m}C_{n}^{*}\left|m,n\right\rangle \left\langle n,m\right|
\]
Here we suppose $C_{n}$ and $C_{m}$ are real, so we have $C_{n}C_{m}^{*}=C_{m}C_{n}^{*}=C_{n}C_{m}$
.

\[
2\rho^{\mathbf{T}}=\sum_{n,m}C_{n}C_{m}\left[\left|n,m\right\rangle \left\langle m,n\right|+\left|m,n\right\rangle \left\langle n,m\right|\right]
\]
we use this formula 
\[
\left|x\right\rangle \left\langle y\right|+\left|y\right\rangle \left\langle x\right|=\frac{\left|x\right\rangle +\left|y\right\rangle }{\sqrt{2}}\frac{\left\langle x\right|+\left\langle y\right|}{\sqrt{2}}-\frac{\left|x\right\rangle -\left|y\right\rangle }{\sqrt{2}}\frac{\left\langle x\right|-\left\langle y\right|}{\sqrt{2}}
\]
$\left|n,m\right\rangle \left\langle m,n\right|+\left|m,n\right\rangle \left\langle n,m\right|$can
be written as
\[
\begin{split} & \left|n,m\right\rangle \left\langle m,n\right|+\left|m,n\right\rangle \left\langle n,m\right|\\
= & \frac{\left|n,m\right\rangle +\left|m,n\right\rangle }{\sqrt{2}}\frac{\left\langle n,m\right|+\left\langle m,n\right|}{\sqrt{2}}-\frac{\left|n,m\right\rangle -\left|m,n\right\rangle }{\sqrt{2}}\frac{\left\langle n,m\right|-\left\langle m,n\right|}{\sqrt{2}}
\end{split}
\]
 Define $\left|a\right\rangle =\frac{\left|n,m\right\rangle +\left|m,n\right\rangle }{\sqrt{2}}$,
$\left|b\right\rangle =\frac{\left|n,m\right\rangle -\left|m,n\right\rangle }{\sqrt{2}}$,
so we have 
\begin{equation}
\left|n,m\right\rangle \left\langle m,n\right|+\left|m,n\right\rangle \left\langle n,m\right|=\left|a\right\rangle \left\langle a\right|-\left|b\right\rangle \left\langle b\right|
\end{equation}
In this way ,we give the diagonal form of $\rho^{\mathbf{T}}$

\begin{equation}
2\rho^{\mathbf{T}}=\sum_{n,m}C_{n}C_{m}\left[\left|a\right\rangle \left\langle a\right|-\left|b\right\rangle \left\langle b\right|\right]
\end{equation}
use this relation 
\[
\sum_{n,m}=\sum_{n=m}+\sum_{n\neq m}
\]
So
\[
2\rho^{\mathbf{T}}=\sum_{n=m}C_{n}C_{m}\left[\left|a\right\rangle \left\langle a\right|-\left|b\right\rangle \left\langle b\right|\right]+\sum_{n\neq m}C_{n}C_{m}\left[\left|a\right\rangle \left\langle a\right|-\left|b\right\rangle \left\langle b\right|\right]
\]
\begin{equation}
2\rho^{\mathbf{T}}=\sum_{n=m}C_{n}C_{m}\left|a\right\rangle \left\langle a\right|+\sum_{n\neq m}C_{n}C_{m}\left[\left|a\right\rangle \left\langle a\right|-\left|b\right\rangle \left\langle b\right|\right]
\end{equation}
\begin{equation}
2\rho^{\mathbf{T}}=\sum_{n=m}C_{n}C_{m}\left|a\right\rangle \left\langle a\right|+2\sum_{n<m}C_{n}C_{m}\left[\left|a\right\rangle \left\langle a\right|-\left|b\right\rangle \left\langle b\right|\right]
\end{equation}
\begin{equation}
\rho^{\mathbf{T}}=\sum_{n=m}C_{n}C_{n}\left|n,n\right\rangle \left\langle n,n\right|+\sum_{n<m}C_{n}C_{m}\left[\left|a\right\rangle \left\langle a\right|-\left|b\right\rangle \left\langle b\right|\right]
\end{equation}
Now it is easy to see the eigenvalues and eigenvectors of $\rho^{\mathbf{T}}$
\begin{equation}
\begin{array}{ccc}
condition & eigenvectors & eigenvalues\\
n=m & \left|n,n\right\rangle  & C_{n}C_{n}\\
n<m & \frac{\left|n,m\right\rangle +\left|m,n\right\rangle }{\sqrt{2}} & +C_{n}C_{m}\\
n<m & \frac{\left|n,m\right\rangle -\left|m,n\right\rangle }{\sqrt{2}} & -C_{n}C_{m}
\end{array}
\end{equation}
In the given $\rho^{\mathbf{T}}$, the $N\left(\rho^{\mathbf{T}}\right)$
is given by 
\begin{equation}
N\left(\rho^{\mathbf{T}}\right)=\sum_{n<m}\left|C_{n}C_{m}\right|=\frac{1}{2}\sum_{n\neq m}\left|C_{n}C_{m}\right|
\end{equation}
So we can get the expression of the entanglement 
\begin{equation}
\begin{split}E_{N}\left(\rho\right) & =\ln\left(1+2\sum_{n<m}\left|C_{n}C_{m}\right|\right)\\
 & =\ln\left(1+\sum_{n\neq m}\left|C_{n}C_{m}\right|\right)\\
 & =\ln\left[\sum_{n}\left|C_{n}\right|^{2}+\sum_{n\neq m}\left|C_{n}C_{m}\right|\right]\\
 & =\ln\left[\sum_{n}\left|C_{n}\right|\right]^{2}\\
 & =2\ln\sum_{n}\left|C_{n}\right|
\end{split}
\end{equation}
with $\sum_{n}\left|C_{n}\right|^{2}=1$. So if we consider
\begin{equation}
\left|\psi_{q}\right\rangle =\sum_{p=0}^{\infty}\sqrt{f_{p}\left(q\right)}\left|p,p+q\right\rangle 
\end{equation}
 
\begin{equation}
\rho=\left|\psi_{q}\right\rangle \left\langle \psi_{q}\right|=\sum_{p_{1}=0}^{\infty}\sum_{,p_{2}=0}^{\infty}\sqrt{f_{p_{1}}\left(q\right)f_{p_{2}}\left(q\right)}\left|p_{1},p_{1}+q\right\rangle \left\langle p_{2},p_{2}+q\right|
\end{equation}
The partial transpose is
\begin{equation}
\rho^{\mathbf{T}}=\sum_{p_{1}=0}^{\infty}\sum_{,p_{2}=0}^{\infty}\sqrt{f_{p_{1}}\left(q\right)f_{p_{2}}\left(q\right)}\left|p_{1},p_{2}+q\right\rangle \left\langle p_{2},p_{1}+q\right|
\end{equation}
it is easy to see the eigenvalues and eigenvectors of $\rho^{\mathbf{T}}$
\begin{equation}
\begin{array}{ccc}
condition & eigenvectors & eigenvalues\\
p_{1}=p_{2} & \left|p_{1},p_{1}+q\right\rangle  & \sqrt{f_{p_{1}}\left(q\right)f_{p_{1}}\left(q\right)}\\
p_{1}<p_{2} & \frac{\left|p_{1},p_{2}+q\right\rangle +\left|p_{2},p_{1}+q\right\rangle }{\sqrt{2}} & +\sqrt{f_{p_{1}}\left(q\right)f_{p_{2}}\left(q\right)}\\
p_{1}<p_{2} & \frac{\left|p_{1},p_{2}+q\right\rangle -\left|p_{2},p_{1}+q\right\rangle }{\sqrt{2}} & -\sqrt{f_{p_{1}}\left(q\right)f_{p_{2}}\left(q\right)}
\end{array}
\end{equation}
the $N\left(\rho^{\mathbf{T}}\right)$ is given by 
\begin{equation}
N\left(\rho^{\mathbf{T}}\right)=\sum_{p_{1}<p_{2}}\left|\sqrt{f_{p_{1}}\left(q\right)f_{p_{2}}\left(q\right)}\right|=\sum_{p_{1}\neq p_{2}}\frac{1}{2}\left|\sqrt{f_{p_{1}}\left(q\right)f_{p_{2}}\left(q\right)}\right|
\end{equation}
So the entanglement of $\rho$ is
\begin{equation}
\begin{split}E_{N}\left(\rho\right) & =\ln\left(1+\sum_{p_{1}\neq p_{2}}\sqrt{f_{p_{1}}\left(q\right)f_{p_{2}}\left(q\right)}\right)\\
 & =\ln\left(\sum_{p_{1}=p_{2}}\sqrt{f_{p_{1}}\left(q\right)f_{p_{2}}\left(q\right)}+\sum_{p_{1}\neq p_{2}}\sqrt{f_{p_{1}}\left(q\right)f_{p_{2}}\left(q\right)}\right)\\
 & =2\ln\sum_{p=0}^{\infty}\sqrt{f_{p}\left(q\right)}
\end{split}
\end{equation}

\subsection*{Supplement[K]: Calculate of $\Omega$}

we have calculate the entanglement 
\begin{equation}
E_{N}^{\mu}\left(q\right)=E_{N}\left(q\right)+\left(q+1\right)\varepsilon\left[\frac{\Omega q+\Omega-1}{2}\varepsilon-1\right]
\end{equation}
with 
\begin{equation}
\Omega=\frac{\sum_{p_{1}=0}^{\infty}\sum_{p_{2}=0}^{\infty}g\left(p_{1},p_{2}\right)}{\left(\sum_{p=0}^{\infty}\sqrt{f_{p}(q)}\right)^{2}}
\end{equation}
\begin{equation}
g\left(p_{1},p_{2}\right)=\frac{f_{p_{2}}\left(q+1\right)f_{p_{1}}\left(q+1\right)}{\sqrt{f_{p_{1}}\left(q\right)f_{p_{2}+1}\left(q\right)}+\sqrt{f_{p_{1}+1}\left(q\right)f_{p_{2}}\left(q\right)}}
\end{equation}
and 
\begin{equation}
f_{p}\left(q\right)=C_{p+q}^{p}\zeta^{p}\left(1-\zeta\right)^{1+q}
\end{equation}
also we have a gaussian approximation for $f_{p}\left(q\right)$,
that is

\begin{equation}
f_{p}\left(q\right)\approx\frac{1}{\sqrt{2\pi}\sigma\left(q\right)}e^{-\frac{\left(p-\kappa\left(q\right)\right)^{2}}{2\sigma\left(q\right)^{2}}}
\end{equation}
with the mean and variance being $\kappa\left(q\right)=\zeta\frac{1+q}{1-\zeta}$
, $\sigma\left(q\right)=\frac{\sqrt{\zeta\left(1+q\right)}}{1-\zeta}$.
If we want to make a simple expression of $E_{N}^{\mu}\left(q\right)$,
we need know the expression of $\Omega$. In perfect measurement,
undering the gaussian approximation:
\begin{equation}
\left(\sum_{p=0}^{\infty}\sqrt{f_{p}\left(q\right)}\right)^{2}\approx\frac{\sqrt{8\pi\zeta\left(1+q\right)}}{1-\zeta}
\end{equation}
First we have a Simplification of $g\left(p_{1},p_{2}\right)$, and
use $f_{p}\left(q\right)=C_{p+q}^{p}\zeta^{p}\left(1-\zeta\right)^{1+q}$
we get 
\[
g\left(p_{1},p_{2}\right)=\sqrt{f_{p_{1}}\left(q\right)f_{p_{2}}\left(q\right)}\frac{\left(1-\zeta\right)^{2}}{\sqrt{\zeta}\left(q+1\right)^{2}}\frac{\left(p_{1}+q+1\right)\left(p_{2}+q+1\right)}{\sqrt{1+\frac{q}{p_{1}+1}}+\sqrt{1+\frac{q}{p_{2}+1}}}
\]
By using gaussian approximation for $f_{p}\left(q\right)\approx\frac{1}{\sqrt{2\pi}\sigma\left(q\right)}e^{-\frac{\left(p-\kappa\left(q\right)\right)^{2}}{2\sigma\left(q\right)^{2}}}$,
we have 
\begin{equation}
\begin{split}g\left(p_{1},p_{2}\right)\approx & \frac{1}{\sqrt{2\pi}\sigma\left(q\right)}\frac{\left(1-\zeta\right)^{2}}{\sqrt{\zeta}\left(q+1\right)^{2}}e^{-\frac{\left(p_{1}-\kappa\left(q\right)\right)^{2}+\left(p_{2}-\kappa\left(q\right)\right)^{2}}{4\sigma{}^{2}}}\\
 & \times\frac{\left(p_{1}+q+1\right)\left(p_{2}+q+1\right)}{\sqrt{1+\frac{q}{p_{1}+1}}+\sqrt{1+\frac{q}{p_{2}+1}}}
\end{split}
\end{equation}
Next we calculation of $\sum_{p_{1}=0}^{\infty}\sum_{p_{2}=0}^{\infty}g\left(p_{1},p_{2}\right)$.
we may calculate $\int_{-\infty}^{+\infty}\int_{-\infty}^{+\infty}g\left(p_{1},p_{2}\right)dp_{1}dp_{2}$
instead of \\ $\sum_{p_{1}=0}^{\infty}\sum_{p_{2}=0}^{\infty}g\left(p_{1},p_{2}\right)$.
So
\begin{equation}
\begin{split} & \int_{-\infty}^{+\infty}\int_{-\infty}^{+\infty}g\left(p_{1},p_{2}\right)dp_{1}dp_{2}\\
= & \frac{1}{\sqrt{2\pi}\sigma\left(q\right)}\frac{\left(1-\zeta\right)^{2}}{\sqrt{\zeta}\left(q+1\right)^{2}}\\
 & \times\int_{-\infty}^{+\infty}\int_{-\infty}^{+\infty}e^{-\frac{\left(p_{1}-\kappa\left(q\right)\right)^{2}+\left(p_{2}-\kappa\left(q\right)\right)^{2}}{4\sigma\left(q\right){}^{2}}}\frac{\left(p_{1}+q+1\right)\left(p_{2}+q+1\right)}{\sqrt{1+\frac{q}{p_{1}+1}}+\sqrt{1+\frac{q}{p_{2}+1}}}dp_{1}dp_{2}
\end{split}
\end{equation}
and we do a taylor expansion for $\frac{\left(p_{1}+q+1\right)\left(p_{2}+q+1\right)}{\sqrt{1+\frac{q}{p_{1}+1}}+\sqrt{1+\frac{q}{p_{2}+1}}}$
:
\[
\begin{split} & \frac{\left(p_{1}+q+1\right)\left(p_{2}+q+1\right)}{\sqrt{1+\frac{q}{p_{1}+1}}+\sqrt{1+\frac{q}{p_{2}+1}}}\\
\approx & \frac{\sqrt{m+1}\left(m+1+q\right)^{\frac{3}{2}}}{2}+\frac{\left(4m+4+q\right)\sqrt{m+1+q}}{8\sqrt{m+1}}\left(p_{1}-m+p_{2}-m\right)
\end{split}
\]
the $m$ is the taylor expansion center. Here we let $m=\kappa\left(q\right)$.
We just calculate zero order with
\begin{equation}
\begin{split} & \int_{-\infty}^{+\infty}\int_{-\infty}^{+\infty}e^{-\frac{\left(p_{1}-\kappa\left(q\right)\right)^{2}+\left(p_{2}-\kappa\left(q\right)\right)^{2}}{4\sigma\left(q\right){}^{2}}}\frac{\sqrt{m+1}\left(m+1+q\right)^{\frac{3}{2}}}{2}dp_{1}dp_{2}\\
= & \frac{\sqrt{m+1}\left(m+1+q\right)^{\frac{3}{2}}}{2}\int_{-\infty}^{+\infty}\int_{-\infty}^{+\infty}e^{-\frac{x^{2}+y^{2}}{4\sigma\left(q\right){}^{2}}}dxdy\\
= & \frac{\sqrt{m+1}\left(m+1+q\right)^{\frac{3}{2}}}{2}4\pi\sigma\left(q\right){}^{2}\\
= & 2\pi\sqrt{m+1}\left(m+1+q\right)^{\frac{3}{2}}\sigma\left(q\right){}^{2}\\
= & 2\pi\sqrt{\kappa\left(q\right)+1}\left(\kappa\left(q\right)+1+q\right)^{\frac{3}{2}}\sigma\left(q\right){}^{2}
\end{split}
\end{equation}
and beacuse $\int_{-\infty}^{+\infty}e^{-x^{2}}xdx=0$, the first
order is 
\begin{equation}
\begin{split} & \int_{-\infty}^{+\infty}\int_{-\infty}^{+\infty}e^{-\frac{\left(p_{1}-\kappa\left(q\right)\right)^{2}+\left(p_{2}-\kappa\left(q\right)\right)^{2}}{4\sigma\left(q\right){}^{2}}}\frac{\left(4m+4+q\right)\sqrt{m+1+q}}{8\sqrt{m+1}}\left(p_{1}-m+p_{2}-m\right)dp_{1}dp_{2}\\
= & 2\int_{-\infty}^{+\infty}\int_{-\infty}^{+\infty}e^{-\frac{\left(p_{1}-\kappa\left(q\right)\right)^{2}+\left(p_{2}-\kappa\left(q\right)\right)^{2}}{4\sigma\left(q\right){}^{2}}}\frac{\left(4m+4+q\right)\sqrt{m+1+q}}{8\sqrt{m+1}}\left(p_{1}-m\right)dp_{1}dp_{2}\\
= & \frac{\left(4m+4+q\right)\sqrt{m+1+q}}{4\sqrt{m+1}}\int_{-\infty}^{+\infty}e^{-\left(\frac{p_{1}-\kappa\left(q\right)}{2\sigma\left(q\right)}\right)^{2}}\left(p_{1}-m\right)dp_{1}\int_{-\infty}^{+\infty}e^{-\left(\frac{p_{2}-\kappa\left(q\right)}{2\sigma\left(q\right)}\right)^{2}}dp_{2}\\
= & \sigma\left(q\right)\sqrt{\pi}\frac{\left(4m+4+q\right)\sqrt{m+1+q}}{2\sqrt{m+1}}\int_{-\infty}^{+\infty}e^{-\left(\frac{p_{1}-\kappa\left(q\right)}{2\sigma\left(q\right)}\right)^{2}}\left(p_{1}-m\right)dp_{1}\\
= & \sigma\left(q\right)^{2}\sqrt{\pi}\frac{\left(4m+4+q\right)\sqrt{m+1+q}}{\sqrt{m+1}}\left[\sigma\left(q\right)\int_{-\infty}^{+\infty}e^{-x^{2}}xdx+\left(\kappa\left(q\right)-m\right)\sqrt{\pi}\right]\\
= & \sigma\left(q\right)^{2}\sqrt{\pi}\frac{\left(4\kappa\left(q\right)+4+q\right)\sqrt{\kappa\left(q\right)+1+q}}{\sqrt{\kappa\left(q\right)+1}}\left[\sigma\left(q\right)\int_{-\infty}^{+\infty}e^{-x^{2}}xdx+\left(\kappa\left(q\right)-\kappa\left(q\right)\right)\sqrt{\pi}\right]\\
= & \sigma\left(q\right)^{3}\sqrt{\pi}\frac{\left(4\kappa\left(q\right)+4+q\right)\sqrt{\kappa\left(q\right)+1+q}}{\sqrt{\kappa\left(q\right)+1}}\left[\int_{-\infty}^{+\infty}e^{-x^{2}}xdx\right]\\
= & 0
\end{split}
\end{equation}
Then we have 
\begin{equation}
\begin{split} & \sum_{p_{1}=0}^{\infty}\sum_{p_{2}=0}^{\infty}g\left(p_{1},p_{2}\right)\\
= & \int_{-\infty}^{+\infty}\int_{-\infty}^{+\infty}g\left(p_{1},p_{2}\right)dp_{1}dp_{2}\\
= & \frac{1}{\sqrt{2\pi}\sigma\left(q\right)}\frac{\left(1-\zeta\right)^{2}}{\sqrt{\zeta}(q+1)^{2}}2\pi\sqrt{\kappa\left(q\right)+1}\left(\kappa\left(q\right)+1+q\right)^{\frac{3}{2}}\sigma\left(q\right)^{2}\\
= & \frac{\sqrt{2\pi}\left(1-\zeta\right)^{2}}{\sqrt{\zeta}(q+1)^{2}}\sqrt{\kappa\left(q\right)+1}\left(\kappa\left(q\right)+1+q\right)^{\frac{3}{2}}\sigma\left(q\right)\\
= & \frac{\sqrt{2\pi}\left(1-\zeta\right)^{2}}{\sqrt{\zeta}(q+1)^{2}}\sqrt{\zeta\frac{1+q}{1-\zeta}+1}\left(\zeta\frac{1+q}{1-\zeta}+1+q\right)^{\frac{3}{2}}\frac{\sqrt{\zeta\left(1+q\right)}}{1-\zeta}
\end{split}
\end{equation}
and 
\begin{equation}
\begin{split}\Omega & =\frac{\sum_{p_{1}=0}^{\infty}\sum_{p_{2}=0}^{\infty}g\left(p_{1},p_{2}\right)}{\left(\sum_{p=0}^{\infty}\sqrt{f_{p}(q)}\right)^{2}}\\
 & \approx\frac{\frac{\sqrt{2\pi}\left(1-\zeta\right)^{2}}{\sqrt{\zeta}(q+1)^{2}}\sqrt{\zeta\frac{1+q}{1-\zeta}+1}\left(\zeta\frac{1+q}{1-\zeta}+1+q\right)^{\frac{3}{2}}\frac{\sqrt{\zeta\left(1+q\right)}}{1-\zeta}}{\frac{\sqrt{8\pi\zeta\left(1+q\right)}}{1-\zeta}}\\
 & =\frac{1}{2}\sqrt{\frac{1+\zeta q}{\zeta+\zeta q}}
\end{split}
\end{equation}
and q is large, we have 
\begin{equation}
\Omega\approx\frac{1}{2}\sqrt{\frac{1+\zeta q}{\zeta+\zeta q}}\approx\frac{1}{2}
\end{equation}
 Then we get 
\begin{equation}
E_{N}^{\mu}\left(q\right)=E_{N}\left(q\right)-\left(q+1\right)\varepsilon+\frac{q^{2}-1}{4}\varepsilon^{2}
\end{equation}

\end{document}